\documentclass[%
 preprint,
 amsmath,amssymb,
 aps,
]{revtex4-2}
\usepackage[english]{babel} 
\usepackage{graphicx}
\graphicspath{{figures/}}
\usepackage{dcolumn}
\usepackage{bm}

\usepackage{subfigure}

\usepackage{amsmath}
\usepackage{tikz}
\usepackage{tikz-cd}
\usetikzlibrary{positioning,arrows} 
\usetikzlibrary{decorations.pathreplacing} 
\usepackage{tikzsymbols} 
\graphicspath{{.},{figures/}}
\usepackage{tikz}
\usetikzlibrary{shapes,arrows}
\usepgflibrary{arrows}
\usetikzlibrary{decorations.pathreplacing}
\usetikzlibrary{spy}
\usetikzlibrary{shapes,arrows}
\usetikzlibrary{calc}
\usetikzlibrary{arrows, arrows.meta}
\usepgflibrary{arrows}

\usetikzlibrary{shapes.geometric, arrows}
\tikzstyle{process} = [rectangle, minimum width=1.5cm, minimum height=1cm, text centered, text width=1.5cm, draw=black, thick, fill=black!30]
\tikzstyle{whiteprocess} = [rectangle, minimum width=1.5cm, minimum height=0.5cm, text centered, text width=2.1cm, draw=black, thick, fill=black!0]
\tikzstyle{decision} = [diamond, minimum width=4cm, minimum height=1cm, text centered, text width=3cm, draw=black, fill=black!20]
\tikzstyle{arrow} = [thick,->,>=stealth]

\usepackage{relsize}

\tikzset{fontscale/.style = {font=\relsize{#1}}
    }
    
\usepackage[export]{adjustbox}

\usepackage{scalerel}
\usepackage{stackengine,wasysym}

\begin{document}


\title[Wavelet based modeling of subgrid-scales in LES of particle-laden turbulent flows]{Wavelet based modeling of subgrid-scales in LES of particle-laden turbulent flows}

\author{M. Hausmann}
\author{F. Evrard}
\altaffiliation[Also at]{ Sibley School of Mechanical and Aerospace Engineering, Cornell University, Ithaca, NY 14853, USA}
\author{B. van Wachem}
\email{berend.vanwachem@ovgu.de}
\affiliation{ 
Chair of Mechanical Process Engineering, Otto-von-Guericke-Universit{\"a}t Magdeburg, \\
  Universit{\"a}tsplatz 2, 39106 Magdeburg, Germany
}

\date{\today}

\begin{abstract}
We propose a novel model to obtain the subgrid-scale velocity in the context of large-eddy simulation (LES) of particle-laden turbulent flows, to recover accurate particle statistics. In the new wavelet enrichment model, the subgrid-scale velocity is discretized with a divergence-free wavelet vector basis, and the coefficients of the expansion are obtained by minimizing the squared error of the linearized subfilter Navier-Stokes equations (SFNSE). The compact support of the wavelet basis is exploited to achieve continuously varying subgrid-scale velocity statistics across the domain. The performance of the new wavelet enrichment model is evaluated in single-phase and particle-laden flow simulations, comparing the results with the results of direct numerical simulations (DNS).  The simulations show that the model can generate inhomogeneous and anisotropic velocity statistics, accurate strain-rotation relations, and a good approximation of the kinetic energy spectrum of the corresponding DNS. Furthermore, the model significantly improves the prediction of the particle-pair dispersion, the clustering of the particles, and the turbulence modulation by particles in two-way coupled simulations. The proposed model recovers the most important interactions between fluid turbulence and the behavior of the particles, while maintaining the computational cost on the order of a LES. 
\end{abstract}

\maketitle

\section{Introduction}
The understanding of the underlying physics of the interactions between turbulence and particles has attracted a lot of research interest for many decades, because of its ubiquity in natural and industrial processes. The only way to capture the most important complex phenomena of these multiphase flows numerically, is to perform direct numerical simulation (DNS), which resolves the turbulent length- and time scales down to the Kolmogorov scales. In single-phase turbulence, extensive research and model development have enabled good predictions of turbulence statistics with computational costs significantly smaller than the costs of a DNS. One of the methods that achieves this, is large eddy simulation (LES), which resolves the large flow structures with the numerical grid while modeling the effect of the smaller, unresolved scales. The success of LES in single-phase turbulence does not apply to particle-laden flows, since various interactions between the phases are not captured by existing LES models. \\
The dispersion and preferential concentration of particles in the turbulent flow can be very different if they are transported with the filtered fluid velocity field instead of the fluid velocity field that contains the full turbulent spectrum \cite{Armenio1999,Ray2011,Fede2006a,Rosa2017}. A solution to this, is to model the unresolved fluid velocity at each particle position. \\
The reconstruction of the subgrid-scale fluid velocity field, even in the absence of particles, has been a topic of interest for decades, which explains the large variety of existing models. The majority of these models, however, suffers from drawbacks that make them unsuitable for the application to particle-laden turbulent flows. \\
A well-known method to generate a fictitious subgrid-scale fluid velocity field, is the kinematic simulation that approximates the fluid velocity field as a truncated Fourier series with coefficients that are chosen from a Gaussian distribution, such that a given kinetic energy spectrum is achieved, and the resulting velocity field is divergence-free \cite{Kraichnan1970,Fung1992,Malik1999}. Since real turbulence is typically not Gaussian and potentially statistically inhomogeneous and anisotropic, the kinematic simulation does not provide a realistic turbulent field. The subgrid-scale velocity field can also be reconstructed by fractal interpolation as proposed by \citet{Scotti1999}. Even though the model is computationally cheap, no physical knowledge of turbulence is incorporated. \citet{Domaradzki1999} proposed a velocity field extrapolation method for the velocity field on a refined grid using explicit filtering of the non-linear advective term in the Navier-Stokes equations. Without further modification, however, the resulting velocity field is not divergence-free. Another family of models either solves the subfilter Navier-Stokes equations (SFNSE) with a simplifying numerical method such as, e.g., partially freezing the velocity field \cite{Terracol2001}, or model the SFNSE itself \cite{Dubrulle2001,Ghate2017,Hughes1998,Hughes2000,Hausmann2022a}. Even though accurate predictions are reported, the models can be computationally expensive or may introduce assumptions that can not be justified in some contexts \cite{Sagaut2005}. \\
Approaches that approximate turbulent velocity fields inspired the development of models for LES of particle-laden flows. A class of models emerged that solve a stochastic differential equation for every particle independently (see, e.g., \cite{Fede2006,Bini2007,Berrouk2007,Shotorban2006,Pozorski2009,Knorps2021}), which turns out to be versatile and computationally efficient. By construction, however, these models predict poor particle pair statistics, and their results strongly depend on the choice of model parameters \cite{Marchioli2017}. Improved particle statistics are observed with models that apply a deconvolution operator on the LES velocity field \cite{Kuerten2006,Shotorban2005,Park2017}, but these models merely modify the velocity field on the LES grid and do not augment the range of modeled scales. \citet{Bassenne2019} combine the dynamical deconvolution of \citet{Park2017} with the subgrid extrapolation of \citet{Domaradzki1999} and obtain realistic particle clustering with LES of homogeneous isotropic turbulence (HIT) and for a wide range of Stokes numbers. The high computational cost, originating from a divergence-free projection on a very fine grid make, this model unsuitable for LES. Recently, \citet{Hausmann2023} propose a model that solves the linearized SFNSE with Fourier basis functions in statistically homogeneous sub-domains and apply it to particle-laden turbulent flows. In LES of HIT, the model mainly recovers particle pair dispersion and clustering of the corresponding DNS for particles of many Stokes numbers. Because of the statistically homogeneous sub-domains, however, the subgrid-scale velocity field is partially discontinuous and requires interpolation. \\
In the present work, we derive a new model for predicting the subgrid-scale fluid velocity field that recovers characteristic properties of turbulence, while maintaining computational costs that are acceptable in the scope of a LES. In order to obtain the right particle statistics and clustering, the modeled subgrid-scale velocity field is expected to have realistic spatial and temporal correlations, strain-rotation relations, and non-Gaussian distributions of, e.g., the fluid velocity gradients. A model for the subgrid-scale velocity must also be able to generate statistically inhomogeneous and anisotropic velocity fields. \\
In our novel model, the SFNSE are linearized by means of the rapid distortion theory and approximately solved by minimizing their squared error. The discrete solution spaces are spanned by divergence-free wavelet vector functions. The narrow support of the wavelet basis in physical and spectral space allows for spatially varying fluid velocity statistics and localization in spectral space, that is required to control the kinetic energy spectrum. Conceptually, the wavelet enrichment is superior to Fourier enrichment, which requires special treatment, such as statistically homogeneous sub-domains, to generate an inhomogeneous subgrid-scale velocity field \cite{Hausmann2022a}. Divergence-free wavelet vector bases are also used by \citet{Letournel2022} to generate a turbulent velocity field by transporting the wavelet coefficients by a stochastic Langevin model. In the newly proposed model, however, a transport equation for the wavelet coefficients is derived that minimizes the error in the momentum balance. \\
Even though the literature on LES of particle-turbulence interaction is dominated by studies and models for the subgrid-scale velocity at the particle positions, to enable one-way coupled simulations, more modeling is required in the case of two-way coupled simulations (i.e., where the turbulence modification by the particles is taken into account). Recently, \citet{Hausmann2023} propose a modeling framework that models the two-way coupling effects in a LES, caused by the unresolved fluid velocity field. \\
The paper is outlined as follows. The novel wavelet enrichment is derived in section \ref{sec:modeling}, and the background of the governing equations and the wavelet basis is provided. In section \ref{sec:simulations}, we introduce the simulation configuration that we use to evaluate the performance of the wavelet enrichment. The results and validation for single-phase and particle-laden flows are given in section \ref{sec:results} before the paper is concluded in section \ref{sec:conclusions}.

\section{Wavelet model for the subgrid-scale velocity}
\label{sec:modeling}
In this section, we describe the novel wavelet enrichment for the subgrid-scale velocity. The subgrid-scale fluid velocity field is projected onto a wavelet basis, and the coefficients of the expansion are determined by locally minimizing the mean squared error of the linearized SFNSE. 

\subsection{Scale decomposition}
LES is based on separating the flow into large-scale contributions and small-scale contributions. This is realized mathematically by filtering a quantity $\varphi$ \cite{Leonard1975}:
\begin{align}
    \tilde{\varphi}(x) = \int\displaylimits_{-\infty}^{\infty}G(x-\xi)\varphi(\xi) \mathrm{d}\xi, 
\end{align}
where $\tilde{.}$ indicates a filtered quantity and $G$ is the filter kernel. Subfilter quantities are given by $\varphi^{\prime} = \varphi - \tilde{\varphi}$, which we also refer to as subgrid-scale quantities in the scope of LES.  Filtering the Navier-Stokes equations (NSE) of an incompressible fluid with constant density $\rho_{\mathrm{f}}$ and kinematic viscosity $\nu_{\mathrm{f}}$ yields the filtered Navier-Stokes equations (FNSE):
\begin{align}
\label{eq:LES1}
    \dfrac{\partial \tilde{u}_i}{\partial x_i} &= 0, \\
\label{eq:LES2}
    \dfrac{\partial \tilde{u}_i}{\partial t} + \tilde{u}_j \dfrac{\partial \tilde{u}_i}{\partial x_j} &= -\dfrac{1}{\rho_{\mathrm{f}}} \dfrac{\partial \tilde{p}}{\partial x_i} + \nu_{\mathrm{f}} \dfrac{\partial^2 \tilde{u}_i}{\partial x_j \partial x_j} - \dfrac{\partial \tau_{ij}}{\partial x_j} + \tilde{s}_i,
\end{align}
where $u_i$ is the velocity and $p$ is the pressure. The source term $s_i$ represents, for instance, the momentum coupling of a second phase or a forcing to maintain statistically stationary turbulence. The subgrid-scale stress tensor accounts for the influence of the subgrid-scales on the filtered scales and is given by
\begin{align}
    \tau_{ij} = \widetilde{u_i u_j} - \tilde{u}_i \tilde{u}_j.
\end{align}
By subtracting the FNSE from the NSE, the governing equations for the subgrid-scale fluid velocity are obtained. In models for the subgrid-scale velocity, the non-linear term in the SFNSE is often omitted, or replaced by, a turbulent viscosity \cite{Dubrulle2001,Laval2001,Hausmann2022a,Ghate2017,Hughes2000}. The resulting linearized SFNSE can be written as
\begin{align}
\label{eq:SmallScales1}
    \dfrac{\partial u_i^{\prime}}{\partial x_i} &= 0, \\
\label{eq:SmallScales2}
    \dfrac{\partial {u}^{\prime}_i}{\partial t} 
    + \tilde{u}_j \dfrac{\partial u^{\prime}_i}{\partial x_j} 
    + u^{\prime}_j \dfrac{\partial \tilde{u}_i}{\partial x_j}&= -\dfrac{1}{\rho_{\mathrm{f}}} \dfrac{\partial p^{\prime}}{\partial x_i} + (\nu_{\mathrm{f}}+\nu_{\mathrm{t}}^{\prime}) \dfrac{\partial^2 u^{\prime}_i}{\partial x_j \partial x_j}
    + \dfrac{\partial \tau_{ij}}{\partial x_j} + s_i^{\prime}.
\end{align}
The turbulent viscosity can be derived from renormalization groups \cite{Canuto1996a}:
\begin{align}
\label{eq:TurbulentViscositySmall}
    \nu_{\mathrm{t}}^{\prime}(k) =  \left(\nu^2_{\mathrm{f}} +\dfrac{2}{5} \int\displaylimits_k^{\infty}q^{-2} E(q) \mathrm{d}q \right)^{1/2} - \nu_{\mathrm{f}},
\end{align}
where $E(k)$ is the kinetic energy spectrum of the wave number $k$. The replacement of the non-linear term with a turbulent viscosity can be justified energetically and by means of their contribution to intermittency \cite{Laval2001}. Similar to subgrid-scale models relying on the Boussinesq hypotheses, however, the linearization mistakenly assumes an alignment of the eigenvectors of the subgrid-scale stress tensor with those of the fluid velocity gradient tensor (see, e.g., Horiuti \cite{Horiuti2003}). Therefore, the use of an eddy viscosity, may it be in the SFNSE or LES subgrid-scale models, are only justified energetically. \\
Similar to our recently proposed model, we aim to efficiently solve the linearized SFNSE to approximate the subgrid-scale fluid velocity field. Instead of expanding the subgrid-scale fluid velocity field in Fourier space \cite{Hausmann2022a}, we approximate it as a finite series of wavelet basis functions. The rapidly decaying support of a wavelet basis in spectral and physical space enables a model that incorporates spectral space information and spatial inhomogeneities.


\subsection{Multiresolution analysis}
\label{ssec:MRA}
\begin{figure}
    \centering
    \subfigure[]{\label{fig:scalingfunctionwaveleta}\includegraphics[scale=0.9]{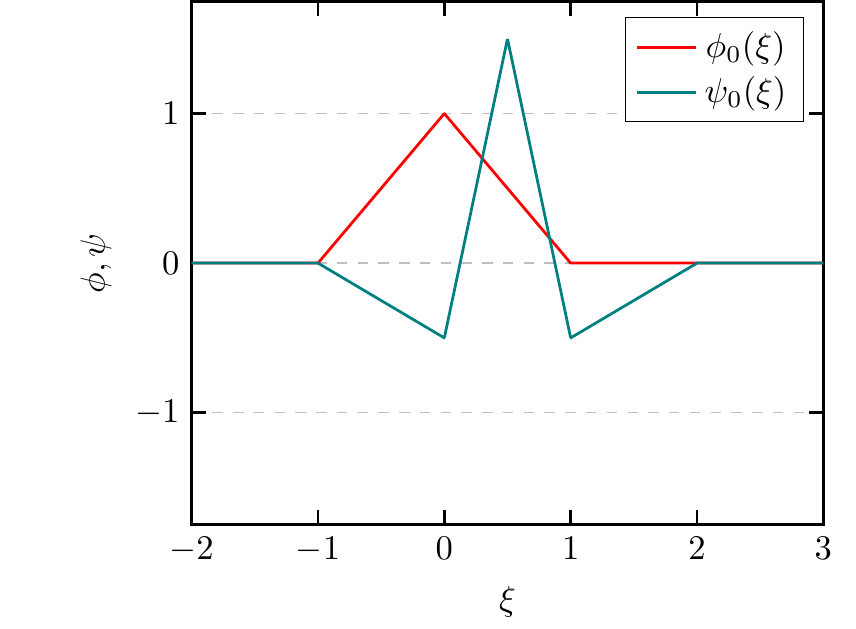}}
    \subfigure[]{\label{fig:scalingfunctionwaveletb}\includegraphics[scale=0.9]{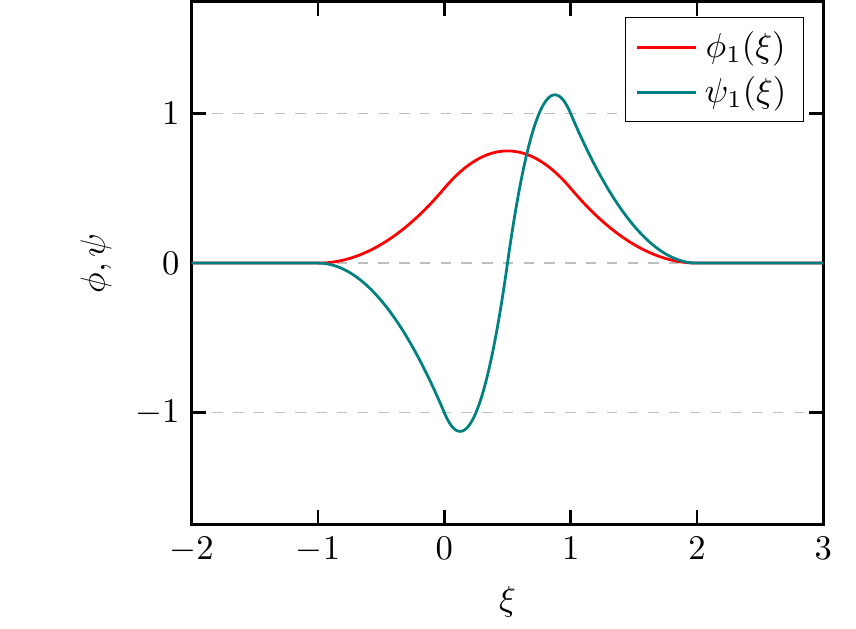}}
    \caption{Plot of the piecewise linear (a) and piecewise quadratic (b) spline scaling and wavelet functions. }
    \label{fig:scalingfunctionwavelet}
\end{figure}

A Fourier series consists of basis functions (i.e., sines and cosines) that are perfectly localized in spectral space (they can be associated to exactly one wave number), but it has no localization in physical space (it is unknown where a frequency occurs). Convoluting the trigonometric basis functions of a Fourier series with a Gaussian window function enables localization of the spectral information in physical space. This is known as the discrete Gabor transform (see, e.g., \cite{Brunton2019a}). Because of the constant window size, the low wave numbers tend to be spectrally underresolved and high wave numbers spatially underresolved. In fact, it is more appropriate to provide a wide physical support for small wave numbers and a narrow physical support for large wave numbers. A multiresolution analysis (MRA) applies this idea \cite{Brunton2019a,Daubechies1992}. A MRA uses scaling functions $\phi$ and wavelets $\psi$ as basis functions, that have compact support (or at least decay rapidly) in physical and spectral space. The scaling functions and the wavelets are elements of function spaces, which have specific properties. We consider sub-spaces of the Lebesgue space $(V_j)_{j\in \mathbb{Z}} \subset L^2(\mathbb{R})$, where the index $j$ can be understood as an indicator of the range of wave numbers the functions possess that are deduced from the respective sub-space. In general, a larger index corresponds to deduced functions of higher wave numbers. The sub-spaces $V_j$ have the following properties \cite{Daubechies1992}:
\begin{itemize}
    \item the sub-spaces are nested $V_j\subset V_{j+1}$,
    \item their intersection is zero $\bigcap_{j\in \mathbb{Z}} V_j = \{ 0\}$,
    \item $\bigcup_{j\in \mathbb{Z}} V_j $ is dense in $L^2(\mathbb{R})$,
    \item they are invariant with respect to scaling $f \in V_j \iff f(2.)\in V_{j+1}$, and
    \item they are invariant with respect to translation $f \in V_j \implies f(.-k)\in V_j, \forall k \in \mathbb{Z} $.
\end{itemize}
Complementary spaces to $V_j$ can be defined such that
\begin{align}
    V_{j+1} = V_j \oplus W_j.
\end{align}
Consequently, we can decompose the $L^2(\mathbb{R})$ as
\begin{align}
    L^2(\mathbb{R}) = V_0 \bigoplus_{j=0}^{\infty}W_j.
\end{align}
From the so-called mother scaling function $\phi \in V_0$ and mother wavelet $\psi \in W_0$ functions are deduced such that $V_j$ is spanned by $\{\phi_{jk}; j,k\in \mathbb{Z} \}$ and $W_j$ is spanned by $\{\psi_{jk}; j,k\in \mathbb{Z} \}$:
\begin{align}
    \phi_{j,k}(\xi) = \phi(2^j\xi-k), \\
    \psi_{j,k}(\xi) = \psi(2^j\xi-k).
\end{align}
Thus, a function $f \in L^2(\mathbb{R})$ may be expressed with a basis of scaling functions and wavelet functions
\begin{align}
    f(\xi) = \displaystyle \sum_{k \in \mathbb{Z}}c_{0,k} \phi_{0,k}(\xi) + \displaystyle \sum_{j\ge 0} \sum_{k \in \mathbb{Z}}d_{j,k} \psi_{j,k}(\xi).
\end{align}
Note that it is common to scale $\phi_{j,k}$ and $\psi_{j,k}$ with a factor of $2^{j/2}$. For simplicity, we absorb this factor in the coefficients $c_{0,k}$ and $d_{j,k}$.
\\
The mother scaling function is defined by the low pass filter
\begin{align}
    \phi\left( \dfrac{\xi}{2} \right) = \displaystyle \sum_{k\in \mathbb{Z}} h_k \phi(\xi-k),
\end{align}
and the mother wavelet function is defined by the high pass filter
\begin{align}
    \psi\left( \dfrac{\xi}{2} \right) = \displaystyle \sum_{k\in \mathbb{Z}} g_k \phi(\xi-k).
\end{align}
The coefficients $h_k,g_k$ characterize the different scaling functions and wavelets with different properties. We express the subgrid-scale velocity as a finite sum of divergence-free wavelet vector basis function as derived by \citet{Lemarie-Rieusset1992}
\begin{align}
\label{eq:expansionsgsvelocity}
    \boldsymbol{u}^{\prime}(\boldsymbol{x}) = \displaystyle \sum_{j=j_{\mathrm{min}}}^{j_{\mathrm{max}}} \sum_{\boldsymbol{k}} \sum_{\epsilon} d_{\mathrm{div},j,\boldsymbol{k}}^{\epsilon} \boldsymbol{\varPsi}_{\mathrm{div},j,\boldsymbol{k}}^{\epsilon}(\boldsymbol{\xi}(\boldsymbol{x})),
\end{align}
where $d_{\mathrm{div},j,\boldsymbol{k}}^{\epsilon}$ are the wavelet coefficients, $\boldsymbol{\varPsi}_{\mathrm{div},j,\boldsymbol{k}}^{\epsilon}$ are divergence-free wavelet vector basis functions, $j_{\mathrm{min}}$ and $j_{\mathrm{max}}$ the limits of the considered levels, and $\boldsymbol{\xi}(\boldsymbol{x})$ is a mapping from physical space coordinates $\boldsymbol{x}$ to the reference coordinates $\boldsymbol{\xi}$. The index $\epsilon$ specifies one of the 14 basis vector functions such that $\boldsymbol{\varPsi}_{\mathrm{div},j,\boldsymbol{k}}^{\epsilon}$ form a Riesz basis of the space of divergence-free vector functions in $\mathbb{R}^3$ \cite{Deriaz2006}. The divergence-free vector functions are composed of four one-dimensional compactly supported spline functions that are depicted in figure \ref{fig:scalingfunctionwavelet} and defined by the coefficients $h_k,g_k$ (in the proposed model similar to coefficients given by Deriaz and Perrier \cite{Deriaz2006} but multiplied with a factor $\sqrt{2}$).\\
The mapping from physical coordinates to reference coordinates is given as
\begin{align}
    \xi_i(x_i) = (x_i - x_{i,\mathrm{min}}) / (x_{i,\mathrm{max}} - x_{i,\mathrm{min}}),
\end{align}
where $x_{i,\mathrm{min}}$ and $x_{i,\mathrm{max}}$ are the limits of the cuboid domain. With the given mapping a periodic basis can be realized for $j\ge3$, which is the first level with a support smaller than one.

\subsection{Local least squares approximation of the model equations}
The subgrid-scale fluid velocity field lies within the space of divergence free vector functions in $\mathbb{R}^3$. Hence, it can be expressed as an infinite series of wavelet vector functions $\boldsymbol{\varPsi}_{\mathrm{div},j,\boldsymbol{k}}^{\epsilon}$ and coefficients $d_{\mathrm{div},j,\boldsymbol{k}}^{\epsilon}$. We are searching coefficients of the finite series expansion Eq. \eqref{eq:expansionsgsvelocity}, such that the subgrid-scale fluid velocity field is approximated well, which is divergence-free and conserves momentum at least by fulfilling the linearized momentum equation \eqref{eq:SmallScales2}. With the expansion of the subgrid-scale velocity with the divergence-free vector functions, the former is satisfied immediately. The latter is achieved by a least squares approximation, which is explained in the following. \\
The advective term in the linearized SFNSE can be Helmholtz decomposed in a divergence-free and a curl-free contribution. Since we seek a solution in the divergence-free wavelet vector space, the curl-free contribution can be dropped together with the pressure term, which gives
\begin{align}
\label{eq:momentumshort}
    \dfrac{\partial {u}^{\prime}_i}{\partial t} + \mathcal{A}_i^{\bot} = \mathcal{D}_i + \mathcal{F}_i^{\bot},
\end{align}
where $\mathcal{A}_i^{\bot} = \left(\tilde{u}_j \dfrac{\partial u^{\prime}_i}{\partial x_j} + u^{\prime}_j \dfrac{\partial \tilde{u}_i}{\partial x_j}\right)^{\bot}$ is an abbreviation for the projected advective term, $\mathcal{D}_i=(\nu_{\mathrm{f}}+\nu_{\mathrm{t}}^{\prime}) \dfrac{\partial^2 u^{\prime}_i}{\partial x_j \partial x_j}$ for the diffusive term, and $\mathcal{F}_i^{\bot}=\left(\dfrac{\partial \tau_{ij}}{\partial x_j}\right)^{\bot}$ for the projected forcing term, which originates from the subgrid-scale stress tensor and mainly supplies the subgrid-scale velocity field with kinetic energy. \\
Minimizing the squared error of equation \eqref{eq:momentumshort} over the whole domain results in the well known Galerkin method, which we refer to as globally optimal solution. In order to make the solution for the coefficients affordable in the scope of a LES, we seek locally optimal solutions for every $j$ and $\boldsymbol{k}$:
\begin{align}
\label{eq:optimizationproblem}
    \dfrac{\partial}{\partial d_{\mathrm{div},j,\boldsymbol{k}}^{\zeta,n+1}} \displaystyle \int \left[\dfrac{\boldsymbol{u}_{j,\boldsymbol{k}}^{\prime,n+1} - \boldsymbol{u}_{j,\boldsymbol{k}}^{\prime,n}}{\Delta t} + \boldsymbol{\mathcal{A}}_{j,\boldsymbol{k}}^{\bot} - \boldsymbol{\mathcal{D}}_{j,\boldsymbol{k}} - \boldsymbol{\mathcal{F}}_{j,\boldsymbol{k}}^{\bot}\right]^2 \mathrm{d}V_x = 0,
\end{align}
which yields the condition
\begin{align}
\label{eq:finalcondition}
    \displaystyle \int \left[\dfrac{\boldsymbol{u}_{j,\boldsymbol{k}}^{\prime,n+1} - \boldsymbol{u}_{j,\boldsymbol{k}}^{\prime,n}}{\Delta t} + \boldsymbol{\mathcal{A}}_{j,\boldsymbol{k}}^{\bot}  - \boldsymbol{\mathcal{D}}_{j,\boldsymbol{k}} - \boldsymbol{\mathcal{F}}_{j,\boldsymbol{k}}^{\bot}\right] \cdot \boldsymbol{\varPsi}_{\mathrm{div},j,\boldsymbol{k}}^{\zeta}(\boldsymbol{\xi}(\boldsymbol{x}))\mathrm{d}V_x = 0.
\end{align}
The fluid velocity field local in $j$ and $\boldsymbol{k}$, that we also refer to as local wave packets, is given as
\begin{align}
    \boldsymbol{u}_{j,\boldsymbol{k}}^{\prime}(\boldsymbol{x}) = \displaystyle \sum_{\epsilon}d_{\mathrm{div},j,\boldsymbol{k}}^{\epsilon} \boldsymbol{\varPsi}_{\mathrm{div},j,\boldsymbol{k}}^{\epsilon}(\boldsymbol{\xi}(\boldsymbol{x})),
\end{align}
and $\boldsymbol{\mathcal{A}}_{j,\boldsymbol{k}}^{\bot}$, $\boldsymbol{\mathcal{D}}_{j,\boldsymbol{k}}$, and $\boldsymbol{\mathcal{F}}_{j,\boldsymbol{k}}^{\bot}$ are the projected advective, diffusive, and projected forcing term, as a function of the local velocity wave packets, respectively. The time derivative is discretized with an explicit Euler-scheme, where $n$ indicates the time level. \\
Relaxing the global optimality condition to local $j,\boldsymbol{k}$ avoids the solution of a large equation system of size $14N_{\mathrm{k}} \times 14N_{\mathrm{k}}$. Instead, we invert $N_{\mathrm{k}}$ systems of size $14\times14$. Note that the matrix is always identical, and in practice only one inversion of the $14\times14$-matrix is required. The linear nature of the projected linearized SFNSE \eqref{eq:momentumshort} allows for superposition of solutions, such as the approximated solutions for the local wave packets $\boldsymbol{u}_{j,\boldsymbol{k}}^{\prime}(\boldsymbol{x})$. If the local wave packets would perfectly satisfy Eq. \eqref{eq:momentumshort}, their superposition would also satisfy Eq. \eqref{eq:momentumshort}. Since $\boldsymbol{u}_{j,\boldsymbol{k}}^{\prime}(\boldsymbol{x})$ is a numerical approximation, the discretization errors add up by superposing the solutions and, in total, lead to an error that is larger than the solutions of the global optimization using the Galerkin method. However, only with the localization of the optimization, and the associated increase of the numerical error compared to the Galerkin method, the solution for the subgrid-scale velocity field becomes computational feasible in the scope of LES. Together with the linearization of the SFNSE, the localization of the optimization constitutes the main assumption of the new wavelet model. \\
Since the wavelet basis possesses no degrees of freedom in the space that is orthogonal to the divergence-free $(L^2(\mathbb{R}^3))^3$, no explicit projection of the advective term is required to obtain a divergence-free subgrid-scale velocity. The projected advective term is obtained by expanding it in a series of wavelet basis functions (similar to the subgrid-scale velocity field) and solving the following problem: 
\begin{align}
    \langle \boldsymbol{\mathcal{A}}_{j,\boldsymbol{k}}^{\bot} | \boldsymbol{\varPsi}_{\mathrm{div},j,\boldsymbol{k}}^{\epsilon}\rangle = \langle \boldsymbol{\mathcal{A}}_{j,\boldsymbol{k}} | \boldsymbol{\varPsi}_{\mathrm{div},j,\boldsymbol{k}}^{\epsilon}\rangle,
\end{align}
where $\langle.|.\rangle$ indicates the inner product. Equation \eqref{eq:finalcondition} is solved with the resulting coefficients, which leads to the same discrete equations as using $\boldsymbol{\mathcal{A}}_{j,\boldsymbol{k}}$ without prior projection in Eq. \eqref{eq:finalcondition}. \\
Note the solution of equation \eqref{eq:finalcondition} is guaranteed to produce the energetically optimal coefficients, because the Hessian
\begin{align}
   \boldsymbol{\mathcal{H}}_{j,\boldsymbol{k}}=\mathcal{H}_{j,\boldsymbol{k}}^{\epsilon,\zeta} &= \dfrac{\partial^2}{\partial d_{\mathrm{div},j,\boldsymbol{k}}^{\epsilon,n+1}\partial d_{\mathrm{div},j,\boldsymbol{k}}^{\zeta,n+1}} \displaystyle \int \left[\dfrac{\boldsymbol{u}_{j,\boldsymbol{k}}^{\prime,n+1} - \boldsymbol{u}_{j,\boldsymbol{k}}^{\prime,n}}{\Delta t} + \boldsymbol{\mathcal{A}}_{j,\boldsymbol{k}}^{\bot} - \boldsymbol{\mathcal{D}}_{j,\boldsymbol{k}} - \boldsymbol{\mathcal{F}}_{j,\boldsymbol{k}}^{\bot}\right]^2 \mathrm{d}V_x \nonumber \\
   &= \dfrac{2}{\Delta t^2}\displaystyle \int \boldsymbol{\varPsi}_{\mathrm{div},j,\boldsymbol{k}}^{\epsilon}(\boldsymbol{\xi}(\boldsymbol{x})) \cdot \boldsymbol{\varPsi}_{\mathrm{div},j,\boldsymbol{k}}^{\zeta}(\boldsymbol{\xi}(\boldsymbol{x}))\mathrm{d}V_x,
\end{align}
is positive-definite, i.e., $\boldsymbol{v}^{\mathrm{T}}\boldsymbol{\mathcal{H}}_{j,\boldsymbol{k}}\boldsymbol{v}>0, \forall \boldsymbol{v}\in \mathbb{R}^{14}\setminus \{ \boldsymbol{0} \}$. It follows, that equation \eqref{eq:optimizationproblem} constitutes a convex optimization problem and the solution has always a minimal error. \\
For a solution of Eq. \eqref{eq:finalcondition} to exist, the following expressions have to be satisfied:
\begin{itemize}
    \item $ \int \displaylimits \boldsymbol{\varPsi}_{\mathrm{div},j,\boldsymbol{k}}^{\epsilon}(\boldsymbol{x}) \cdot \boldsymbol{\varPsi}_{\mathrm{div},j,\boldsymbol{k}}^{\zeta}(\boldsymbol{x}) \mathrm{d}V_x<\infty,$
    \item $\int \displaylimits \dfrac{\partial \left(\boldsymbol{\varPsi}_{\mathrm{div},j,\boldsymbol{k}}^{\epsilon}(\boldsymbol{x}) \cdot \boldsymbol{\varPsi}_{\mathrm{div},j,\boldsymbol{k}}^{\zeta}(\boldsymbol{x})\right) }{\partial x_j} \mathrm{d}V_x<\infty,$
    \item $\int \displaylimits \boldsymbol{\varPsi}_{\mathrm{div},j,\boldsymbol{k}}^{\epsilon}(\boldsymbol{x}) \cdot \nabla^2\boldsymbol{\varPsi}_{\mathrm{div},j,\boldsymbol{k}}^{\zeta}(\boldsymbol{x}) \mathrm{d}V_x<\infty,$
\end{itemize}
where $\boldsymbol{\varPsi}_{\mathrm{div},j,\boldsymbol{k}}^{\epsilon}$ must possess second order weak derivatives. The divergence-free basis introduced in section 
\ref{ssec:MRA} satisfies these requirements.


\subsection{Realization of the model in practice}
\label{ssec:realization}
The forcing term $\mathcal{F}_i$ represents the energy transfer from the resolved scales to the subgrid-scales and is modeled in the proposed framework. A direct evaluation of $\mathcal{F}_i=\dfrac{\partial \tau_{ij}}{\partial x_j}$ requires explicit filtering with an a priori unknown filter. Kinetic energy can be added to the subgrid-scale velocity by the advective terms and the subgrid-scale stress tensor, that both underlie modeling assumptions (e.g., the linearization and the explicit filtering) and discretization errors. Consequently, undesired high or low kinetic energies can occur and even destabilize the numerical solution. \\
Since it is essential for the subgrid-scale velocity to have a realistic kinetic energy, we seek a model forcing term that can easily adjust the kinetic energy in both directions. The advective term $\mathcal{A}_i$ can add and remove kinetic energy across the scales. The diffusive term $\mathcal{D}_i$, however, always removes kinetic energy if the viscosity is positive. The advective term and the diffusive term are scaled to remove kinetic energy from or add kinetic energy to the subgrid-scale velocity field. The total viscosity is replaced by an effective viscosity:
\begin{align}
    \nu_{\mathrm{eff}}(k) =  \sqrt{\dfrac{K_{j,\boldsymbol{k}}}{K_{j,\mathrm{desired}}} }\left( \nu_{\mathrm{f}}+\nu_{\mathrm{t}}^{\prime}(k) \right).
\end{align}
The kinetic energy of each wavelet is obtained with
\begin{align}
\label{eq:actualKE}
    K_{j,\boldsymbol{k}} = \dfrac{C_{d\rightarrow K}}{N_{\epsilon}}\displaystyle \sum_{\epsilon} (d_{\mathrm{div},j,\boldsymbol{k}}^{\epsilon})^2,
\end{align}
where $C_{d\rightarrow K}$ is a constant that depends on the distribution of $d_{\mathrm{div},j,\boldsymbol{k}}^{\epsilon}$ and is frequently adjusted by rearranging Eq. \eqref{eq:actualKE} and sampling $K_{j,\boldsymbol{k}}=(\boldsymbol{u}_{j,\boldsymbol{k}}^{\prime}\cdot\boldsymbol{u}_{j,\boldsymbol{k}}^{\prime})/2$ at random positions. The desired kinetic energy follows the scaling of the spectrum of the inertial range
\begin{align}
    K_{j,\mathrm{desired}} = \dfrac{K_{\mathrm{desired}}}{\mathcal{N}}k_j^{-5/3}\Delta k_j,
\end{align}
with the norm
\begin{align}
    \mathcal{N} = \displaystyle \sum_j k_j^{-5/3} \Delta k_j,
\end{align}
and the wave number $k_j = 2^j2\pi/L$. The wave number step is given by
\begin{align}
    \Delta k_j = \dfrac{k_{j+1}-k_{j-1}}{2},
\end{align}
and $L$ indicates the size of the domain. Since the basis functions are associated with a range of wave numbers, $k_j$ must be interpreted as a characteristic wave number of the respective level $j$. \\ 
The advective term is also scaled with a factor of $\sqrt{ K_{j,\mathrm{desired}}/K_{j,\boldsymbol{k}}}$, which accelerates the adaption to the desired kinetic energy. This procedure for maintaining the kinetic energy is stable and insensitive to external disturbances, such as momentum sources originating from two-way coupled particles. Note that no forcing is introduced that prescribes a specific (typically Gaussian) probability distribution function (PDF). \\
As described in section \ref{ssec:MRA}, the wavelet basis functions are shifted by discrete integer values in space. This leads to points in space that coincide with peaks of the basis functions and, consequently, to statistically higher kinetic energy at these points. In order to avoid such a spatial bias, the basis functions are not fixed in space but move with a velocity much smaller than the eddy-turnover time of the respective level and within a small region near their original position. There are many ways of realizing a random movement of the wavelets. The method we applied is described in the Appendix \ref{sec:appendix}. Note that the movement of the wavelets solely requires modification of the sampling of the subgrid-scale velocity because the coefficients $d_{\mathrm{div},j,\boldsymbol{k}}^{\epsilon}$ are determined independently for every $j$ and $\boldsymbol{k}$. 

\section{Simulation setups}
\label{sec:simulations}
The predictions of the proposed wavelet enrichment model are evaluated by means of different single-phase and particle-laden flow configurations. In this section the numerical solution of the flow and the particle transport is briefly introduced, and the parameters of the simulations are provided.

\subsection{Numerical solution}
\begin{figure}[h]
    \centering
    \includegraphics[scale=1.5]{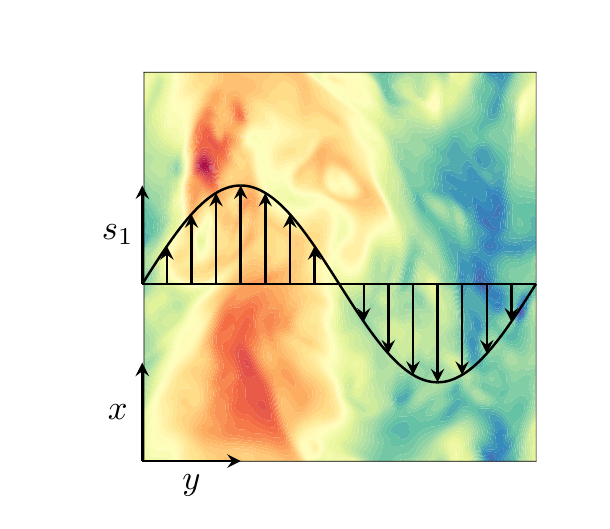}
    \caption{Sketch of the turbulent shear flow configuration. A source term in x-direction is added that varies with one period of a sin-profile across the y-direction and drives a turbulent shear flow. The black arrows indicate the profile of the momentum source and the color represents a slice of the velocity in x-direction.}
    \label{fig:shearflow}
\end{figure}

The NSE and FNSE are solved numerically for different configurations in cubic domains of size $L$ and periodic boundary conditions in every directions. The subgrid-scale stress tensor in the LES is modeled with the localized dynamic kinetic energy model (LDKM) as proposed by Menon and coworkers \cite{Kim1999, Menon1996} and extended to two-way coupled particle-laden flows by \citet{Hausmann2023} (referred to as mLDKM). Note that the mLDKM requires a model for the subgrid-scale velocity at the positions of the particles to be closed. A second order finite volume solver is used to numerically solve the flow equations. More details on the flow solver may be found in \citet{Denner2020} and \citet{Bartholomew2018}. \\
The investigated configurations differ in the source terms that appear in the momentum equations. We consider simulations of HIT, where a statistically steady state is obtained by continuously supplying kinetic energy through turbulence forcing. The forcing procedure is described in \citet{Mallouppas2013a}. \\
A second type of flow which is investigated in this paper, is a turbulent shear flow. This flow is simulated by a constant momentum source that varies as a sin-function across the domain:
\begin{align}
    \boldsymbol{s} = s_{\mathrm{max}} \sin (2\pi y/L)\boldsymbol{e}_1,
\end{align}
where $s_{\mathrm{max}}$ is the amplitude of the momentum source, and $\boldsymbol{e}_1$ is the basis vector in the x-direction. For this turbulent shear flow configuration a shear Reynolds number is defined $Re_{\mathrm{shear}}=L\sqrt{s_{\mathrm{max}}L}/\nu_{\mathrm{f}}$. A sketch of this configuration is provided in figure \ref{fig:shearflow}. \\
We also assess the wavelet enrichment model using a simulation configuration with two-way coupled point-particles, where the particle-induced momentum source is taken into account with the particle-source-in-cell (PSIC) method \cite{Crowe1977}
\begin{align}
    \boldsymbol{s} = -\dfrac{1}{\rho_{\mathrm{f}} V_{\mathrm{cell}}} \sum_{p \in \Omega_{\mathrm{cell}}}\boldsymbol{F}_{p},
\end{align}
where $\Omega_{\mathrm{cell}}$ represents a computational grid cell used in the finite volume solver that has a volume of $V_{\mathrm{cell}}$. The particles are treated in a Lagrangian framework, and their motion is governed by the fluid-particle interface force, $\boldsymbol{F}_p$. We consider small, heavy, and  spherical particles that we assumed to be only influenced by the drag force $\boldsymbol{F}_p=\boldsymbol{F}_{\mathrm{D},p}$, which we compute using Stokes' law augmented with the Schiller-Naumann correlation \cite{Schiller1933}
\begin{align}
   \boldsymbol{F}_{\mathrm{D},p} = C_{\mathrm{D}} \dfrac{\rho_{\mathrm{f}}}{8} \pi d_p^2 |\boldsymbol{u}_{\mathrm{rel}}| \boldsymbol{u}_{\mathrm{rel}},
\end{align}
with the drag coefficient
\begin{align}
    C_{\mathrm{D}} = \dfrac{24}{Re_p}(1+0{.}15 Re_p^{0{.}687}), 
\end{align}
where $d_p$ is the diameter of the particle with the index $p$ and $Re_p=\boldsymbol{u}_{\mathrm{rel}}d_p/\nu_{\mathrm{f}}$ is its Reynolds number. The relative velocity is the difference between the velocity of the particle $\boldsymbol{v}_p$ and the fluid velocity $\boldsymbol{u}(\boldsymbol{x}_p)$ at the position of the particle $\boldsymbol{x}_p$:
\begin{align}
    \boldsymbol{u}_{\mathrm{rel}} = \boldsymbol{u}(\boldsymbol{x}_p) - \boldsymbol{v}_p.
\end{align}
The particle position changes according to
\begin{align}
\label{eq:particleeom1}
    \dfrac{\mathrm{d} \boldsymbol{x}_p}{\mathrm{d} t} = \boldsymbol{v}_p,
\end{align}
and the particle velocity according to Newton's second law
\begin{align}
\label{eq:particleeom2}
    \dfrac{\mathrm{d} \boldsymbol{v}_p}{\mathrm{d} t} = \dfrac{1}{\rho_p V_p} \boldsymbol{F}_p,
\end{align}
where $\rho_p$ is the density of the particle. The motion of the particles is obtained using the Verlet-scheme \cite{Verlet1967}. The fluid velocity is interpolated to the particle position by divergence-free interpolation \cite{Toth2002}.


\subsection{Parameters of the simulations}
We consider four different simulation configurations to assess the predictions of the proposed model: forced HIT of a single-phase flow (HIT-f-s), a single-phase turbulent shear flow (TS-s), forced HIT with one-way coupled particles of five different Stokes numbers (HIT-f-1wc), and decaying HIT with two-way coupled particles of Stokes number $St=8$ and a particle mass fraction of $\phi_{\mathrm{m}}=1$ (HIT-d-2wc). In table \ref{tab:configurations}, important parameters of the four simulations are summarized. The flow quantities of the simulation HIT-d-2wc are given for the corresponding single-phase flow and before the onset of the decay of the turbulence. \\
The forcing is only applied in the wave number range $kL/2\pi\in[3,6]$ for the forced HIT simulations. For the HIT simulations, the resolution is $k_{\mathrm{max}}\eta = 1{.}37$, and for the turbulent shear flow it is $k_{\mathrm{max}}\eta = 1{.}16$. The maximum resolved wave number is defined as $k_{\mathrm{max}}= \pi N_{\mathrm{DNS}}/L$. \\
In the LES of the HIT with the wavelet enrichment the range of considered levels is $j\in [4,6]$ and in the LES of the turbulent shear flow $j\in [3,5]$. The lowest levels are chosen such that their characteristic wave numbers correspond to the respective cutoff wave numbers of the LES. The upper level limits are chosen such that the significant amount of kinetic energy is captured with the modeled wave numbers. \\

\begin{table}[htb]
\caption{\label{tab:configurations}Summary of the parameters that define the four simulations configurations.}
\begin{ruledtabular}
\begin{tabular}{ccccc}
Name & HIT-f-s & TS-s & HIT-f-1wc & HIT-d-2wc\\
\hline
Flow type & forced HIT & turb. shear flow & forced HIT & decaying HIT \\
Energy supply & turb. forcing & sinusoidal force & turb. forcing & - \\
Particles & - & - & one-way coupled & two-way coupled \\
Taylor Reynolds number $Re_{\lambda}$ & $75$ & - & $75$ & $75$ \\
Turbulent Reynolds number $Re_{\mathrm{l}}$ & $205$ & - & $205$ & $205$  \\
Shear Reynolds number $Re_{\mathrm{shear}}$ & - & $3115$ & - & - \\
Reference time $T_{\mathrm{ref}}$ & $L/\sqrt{2/3\langle K\rangle}$ & $\sqrt{L/s_{\mathrm{max}}}$ & $L/\sqrt{2/3\langle K\rangle}$ & $L/\sqrt{2/3\langle K\rangle}$ \\
Kolmogorov length scale $\eta/L$ & $0{.}0017$ & $0{.}0029$ & $0{.}0017$ & $0{.}0017$ \\
Kolmogorov time scale $\tau_{\eta}/T_{\mathrm{ref}}$ & $0{.}0075$ & $0{.}026$ & $0{.}0075$ & $0{.}0075$ \\
Computational cells DNS $N_{\mathrm{DNS}}^3$ & $256^3$ & $128^3$ & $256^3$ & $256^3$ \\
Computational cells LES $N_{\mathrm{LES}}^3$ & $32^3$ & $16^3$ & $32^3$ & $32^3$ \\
Number of particles $N_{\mathrm{p}}$ & - & - & $5\times480115$ & $12057066$ \\
Stokes number $St$ & - & - & $0{.}5,1,2,4,8$ & $8$ \\
\end{tabular}
\end{ruledtabular}
\end{table}
\section{Results and Discussions}
\label{sec:results}
In the present section we evaluate the predictions of the wavelet enriched LES (LES-WL) and compare it to a LES without enrichment and the corresponding DNS. We separately discuss the single-phase flow statistics and the statistics of particle-laden flows. 

\subsection{Single-phase flow statistics}
\begin{figure}[htp]
    \centering
    \subfigure[]{\label{fig:contoura}\includegraphics[scale=0.8]{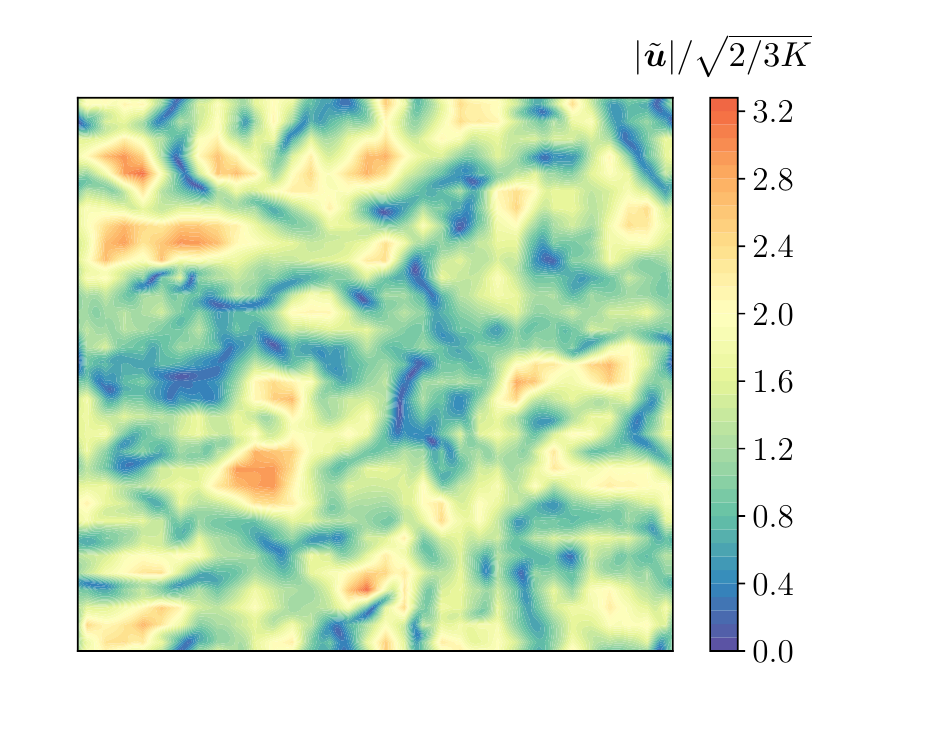}}
    \subfigure[]{\label{fig:contourb}\includegraphics[scale=0.8]{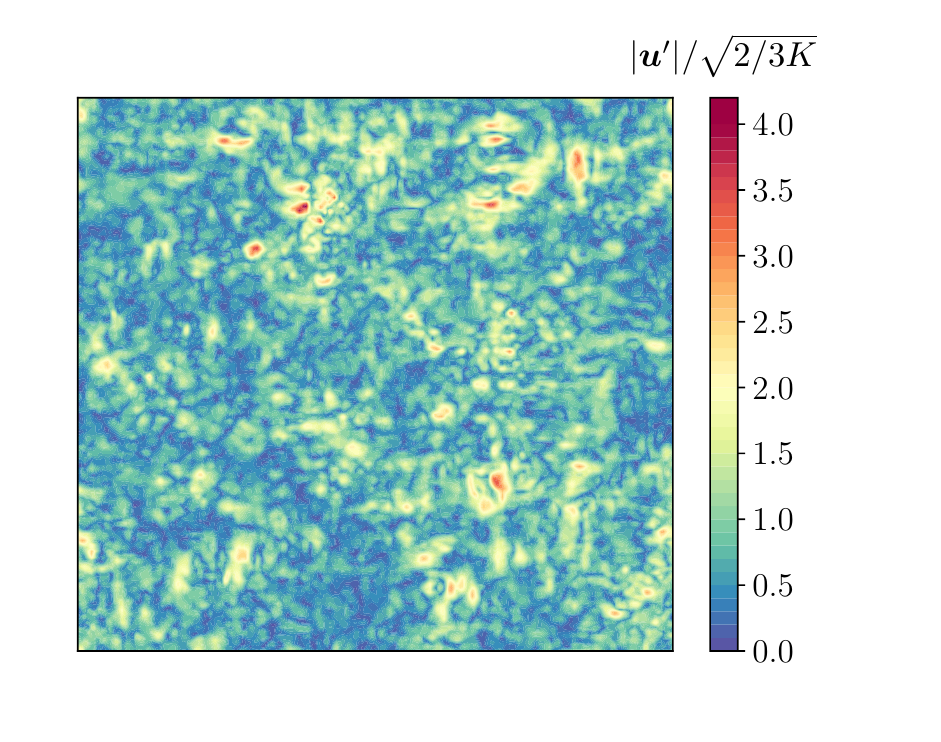}}
    \subfigure[]{\label{fig:contourc}\includegraphics[scale=0.8]{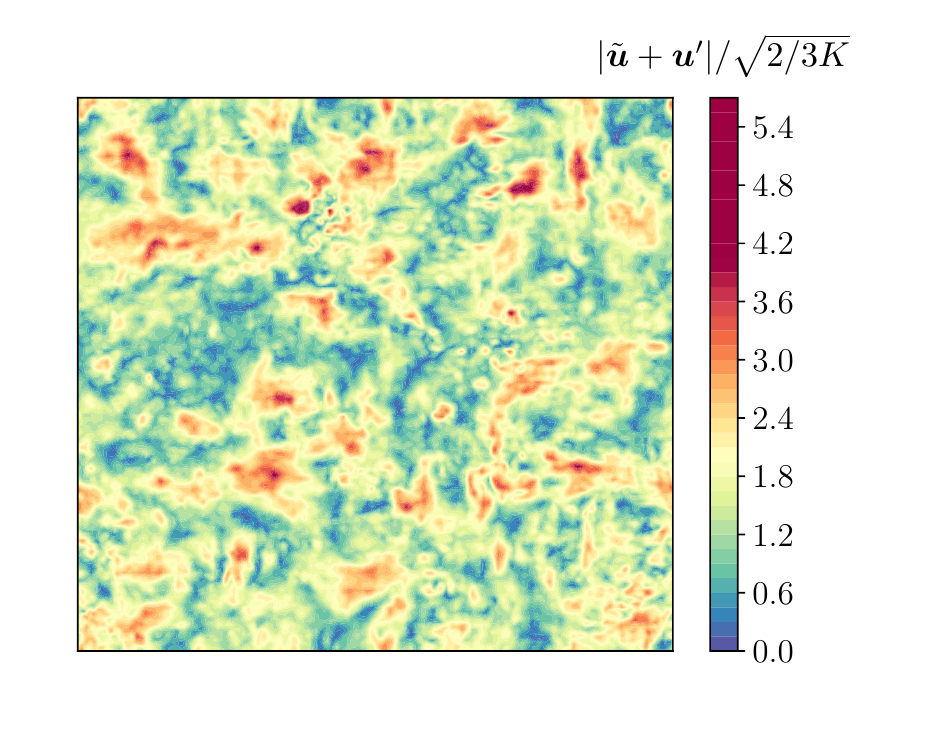}}
    \subfigure[]{\label{fig:contourd}\includegraphics[scale=0.8]{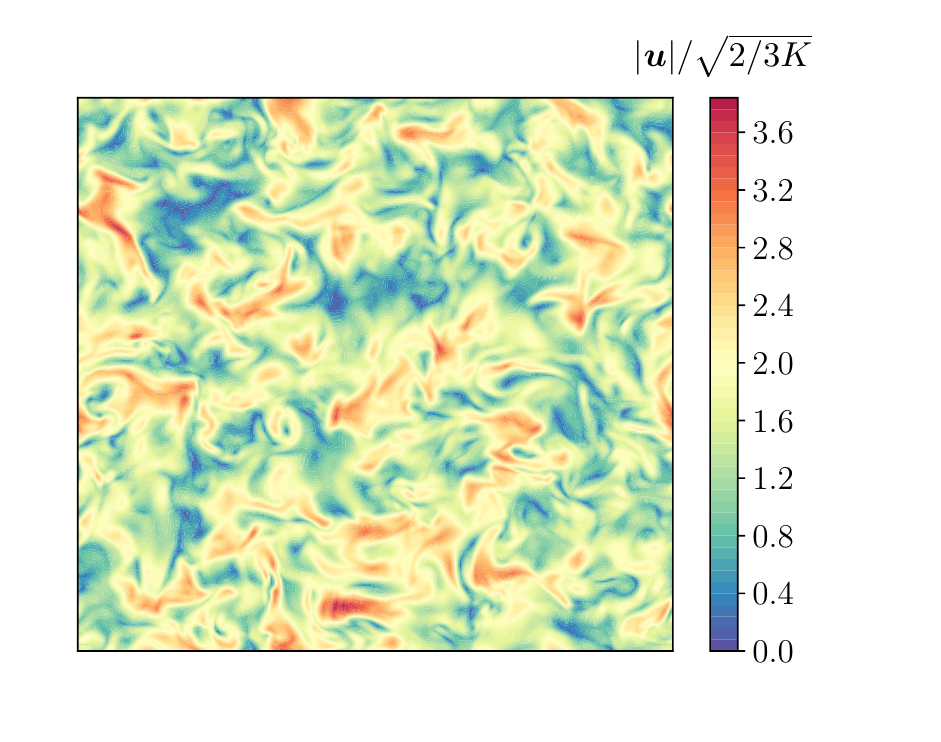}}
    \caption{Slice of the velocity magnitude for the simulation configuration HIT-f-s. The results are shown for the LES (a), the subgrid-scale velocity obtained from the wavelet enrichment (b), the superposition of the LES velocity and the the modeled subgrid-scale velocity (c), and the DNS (d). The color scaling of all subfigures is the same even though the maximum velocities are different. }
    \label{fig:contour}
\end{figure}

\begin{figure} [htp]
    \centering
    \subfigure[]{\label{fig:spectrumQa}\includegraphics[scale=0.9]{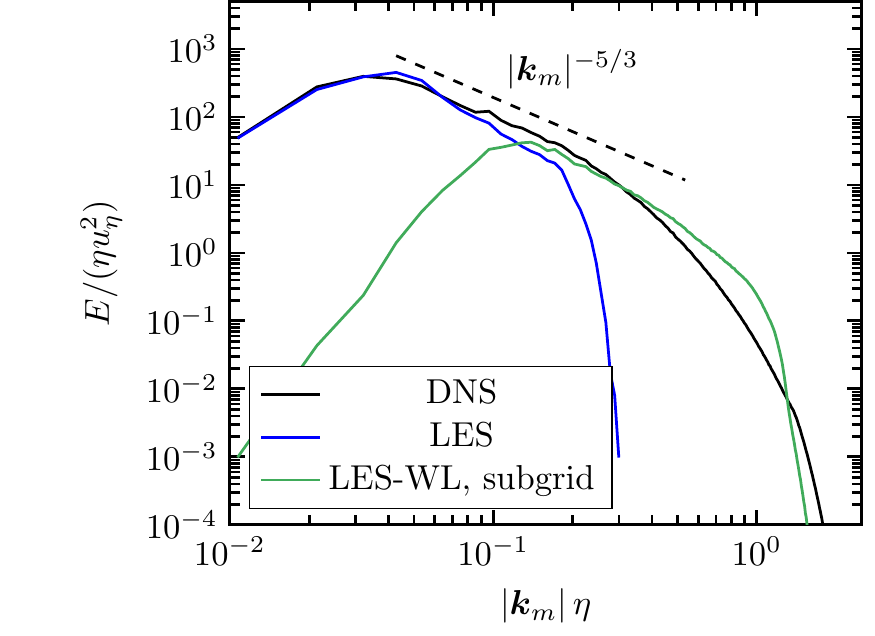}}
    \subfigure[]{\label{fig:spectrumQb}\includegraphics[scale=0.9]{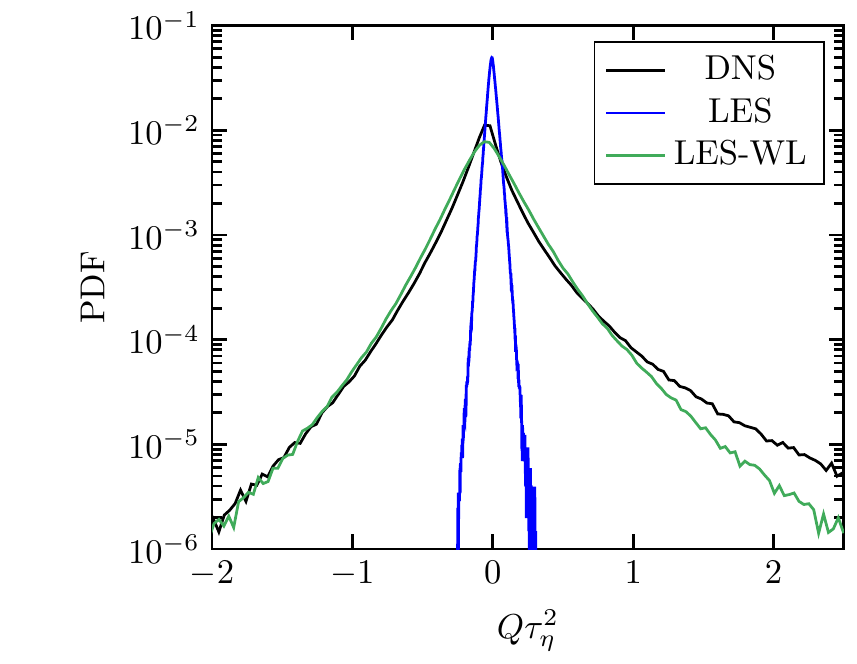}}
    \caption{Kinetic energy spectrum (a) and PDF of the second invariant of the velocity gradient tensor (b) for the simulation configuration HIT-f-s. The results are shown for the DNS, the LES, and the wavelet enriched LES (LES-WL). The inertial range slope is plotted for comparison. }
    \label{fig:spectrumQ}
\end{figure}

We first consider the flow statistics of the forced HIT configuration (HIT-f-s). Figure \ref{fig:contour} compares the normalized fluid velocity magnitudes of the LES, the modeled subgrid-scale velocity obtained from the wavelet enrichment, and their sum with the DNS. The small flow structures, which are absent in the LES and provided by the wavelet enrichment, possess regions with different kinetic energies. This is possible because the compact support of the wavelet basis enables varying statistics across the domain. Some small regions with very high kinetic energy are observed. Such rare high intensity events are characteristic for turbulence and an indicator for intermittency. The superposition of the LES velocity and the modeled subgrid-scale velocity approximates the DNS velocity, which is widely similar in magnitude and range of length scales, but shows differences in the shape of the flow structures. The wavelet enrichment minimizes the local errors and, hence, prevents the systematical formation of flow structures with the neighbor wave packets. \\
The similarities of the velocity fields of the DNS and the wavelet enriched LES can be quantified by comparing the kinetic energy spectra $E$ normalized by the Kolmogorov length scale $\eta$ and velocity $u_{\eta}$ as shown in figure \ref{fig:spectrumQ}. The kinetic energy of the small wave numbers is recovered well by the LES. For the unresolved fluid velocity field the LES, the kinetic energy spectrum of the modeled subgrid-scale fluid velocity field approximates the kinetic energy spectrum of the DNS. Except for an overprediction of the spectrum of relatively small energy at high wave numbers, the general trend is captured. The observed deviation originates from the inertial range power law scaling prescribed by the forcing and may be improved if knowledge of the dissipation range is incorporated in the model. \\
As shown analytically by \citet{Maxey1987}, the clustering of inertial particles of very small Stokes number is proportional to the second invariant of the fluid velocity gradient tensor
\begin{align}
    Q = \dfrac{1}{2}\left( \Omega_{ij}\Omega_{ij} - S_{ij}S_{ij} \right),
\end{align}
with the rotation-rate tensor $\Omega_{ij}$ and the strain-rate tensor $S_{ij}$. Therefore, to be able to correctly predict clustering, an accurate prediction of this tensor is crucial. Figure \ref{fig:spectrumQ} shows the PDF of the second invariant of the velocity gradient tensor for the LES, the wavelet enriched LES and the DNS. It can be observed that the LES by itself does not predict very high strain and low vorticity, or high vorticity and low strain events as is predicted by the DNS. With the wavelet enriched LES, such events occur, even with a very similar probability as in the DNS; only the degree of asymmetry of the PDF (i.e., the higher probability of high rotation low strain events) is slightly underestimated as compared to the DNS. With the linearization of the SFNSE, the non-linear term, as a contribution to the second invariant of the velocity gradient tensor, is modeled by a turbulent viscosity,
that does not affect the second invariant of the velocity gradient tensor the same way as the non-linear term. This linearization can be an explanation for the observed deviation between the wavelet enriched LES and the DNS. \\

\begin{figure}
    \centering
    \subfigure[]{\label{fig:autcorrelationsa}\includegraphics[scale=0.9]{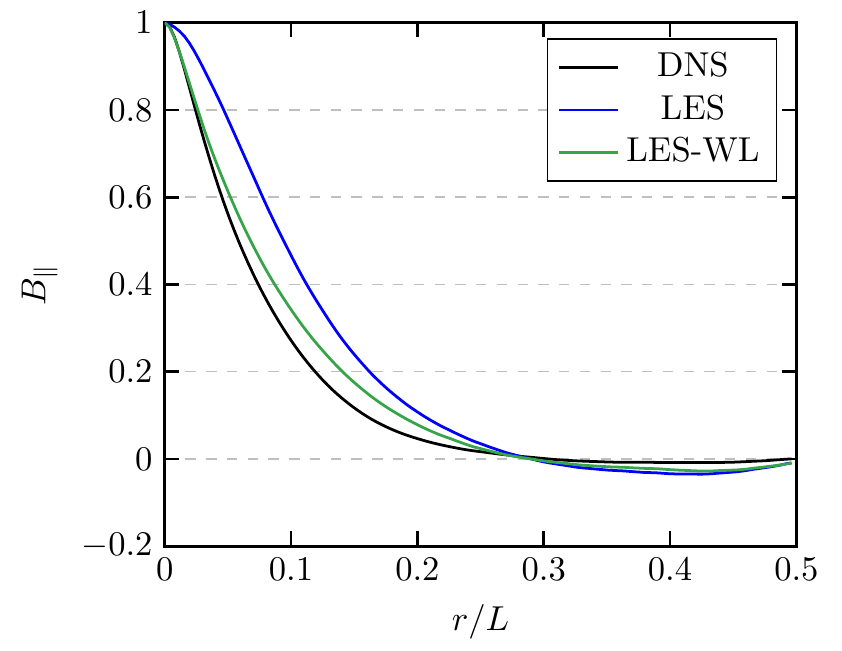}}
    \subfigure[]{\label{fig:autocorrelationsb}\includegraphics[scale=0.9]{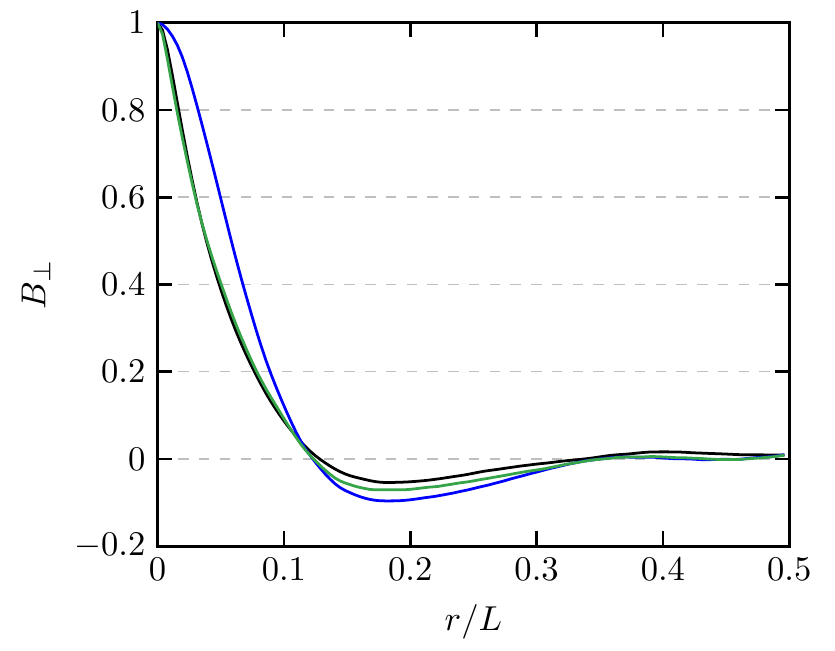}}
    \caption{Longitudinal (a) and transverse (b) fluid velocity autocorrelations over the distance $r$ for the simulation configuration HIT-f-s. The results are shown for the DNS, the LES, and the wavelet enriched LES (LES-WL). }
    \label{fig:autcorrelations}
\end{figure}

\begin{figure}[htp]
    \centering
    \subfigure[]{\label{fig:gradientsa}\includegraphics[scale=0.9]{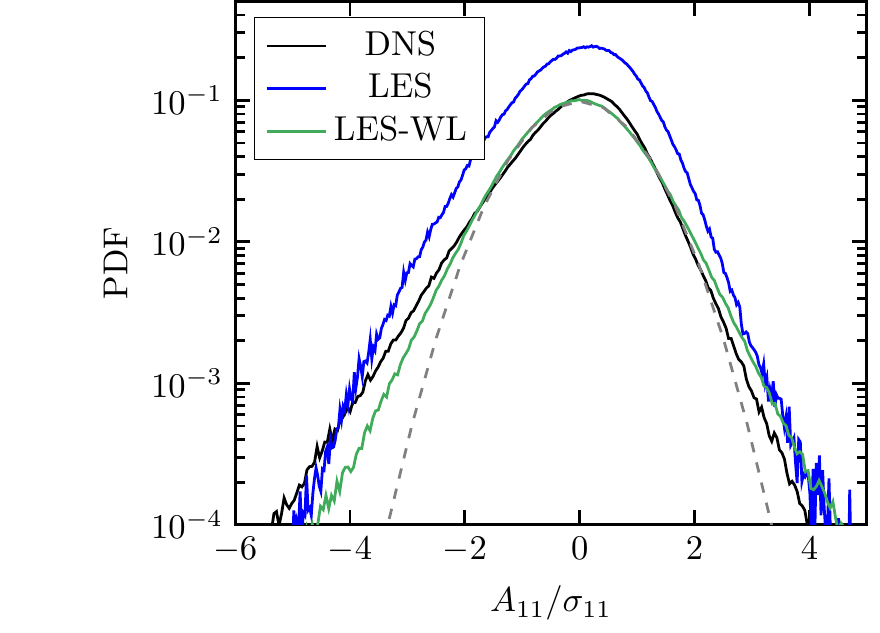}}
    \subfigure[]{\label{fig:gradientsb}\includegraphics[scale=0.9]{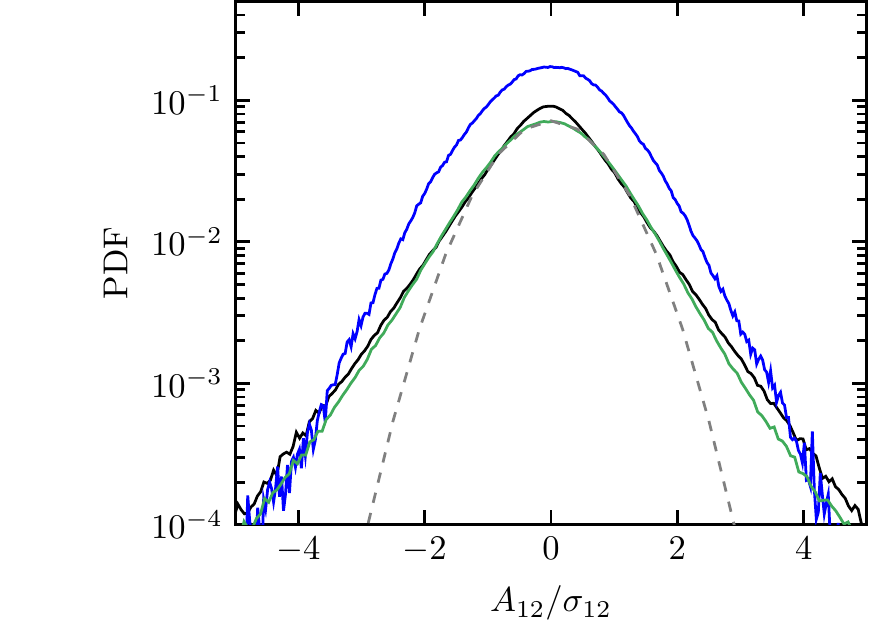}}
    \caption{PDF of longitudinal (a) and transverse (b) velocity gradients normalized by their standard deviations for the simulation configuration HIT-f-s. The results are shown for the DNS, the LES, and the wavelet enriched LES (LES-WL). The dashed lines represent a Gaussian.}
    \label{fig:gradients}
\end{figure}

\begin{figure}
    \centering
    \subfigure[]{\label{fig:shearflowcorrelationsa}\includegraphics[scale=0.9]{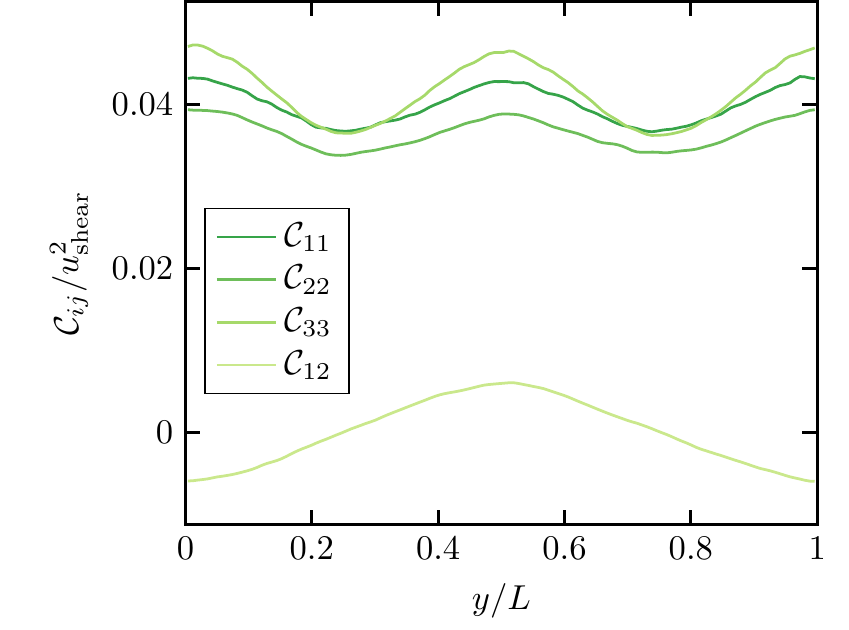}}
    \subfigure[]{\label{fig:shearflowcorrelationsb}\includegraphics[scale=0.9]{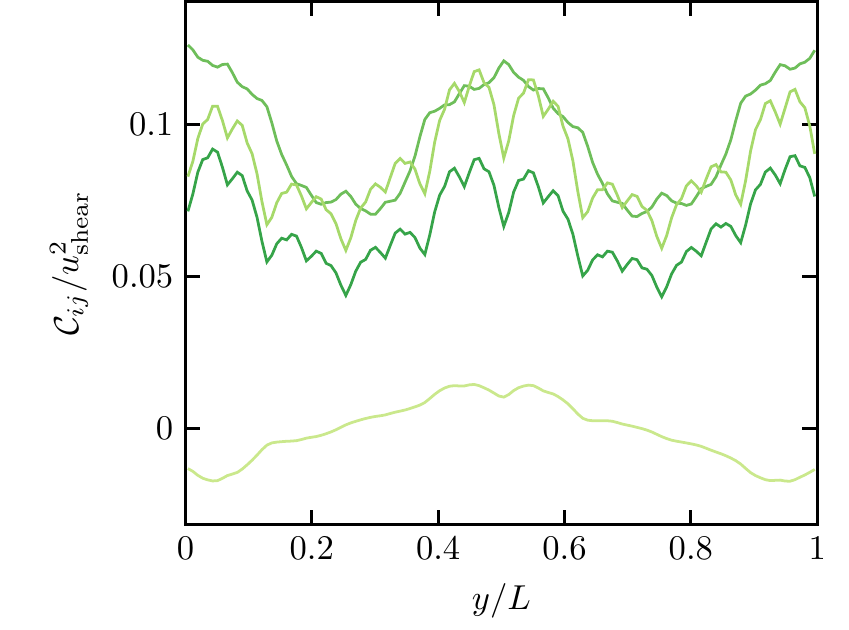}}
    \caption{Normalized spatial correlations in the y-direction for the simulation configuration TS-s. (a) the correlations of the DNS explicitly filtered with a spectrally sharp filter, and (b) the correlations of the subgrid-scale velocity that is modeled with the wavelet enrichment.}
    \label{fig:shearflowcorrelations}
\end{figure}

Figure \ref{fig:autcorrelations} shows the longitudinal and transverse fluid velocity autocorrelation functions, $B_{\parallel}$ and $B_{\perp}$, for the DNS, the LES without wavelet enrichment and the wavelet enriched LES. For homogeneous isotropic flows, they are defined as \cite{George2010}
\begin{align}
    B_{\parallel}(r) = \dfrac{\langle u_{\alpha}(\boldsymbol{x},t) u_{\alpha}(\boldsymbol{x}+r\boldsymbol{e}_{\parallel},t)\rangle}{\langle u_{\alpha}(\boldsymbol{x},t) u_{\alpha}(\boldsymbol{x},t)\rangle}, \\
    B_{\perp}(r) = \dfrac{\langle u_{\alpha}(\boldsymbol{x},t) u_{\alpha}(\boldsymbol{x}+r\boldsymbol{e}_{\perp},t)\rangle}{\langle u_{\alpha}(\boldsymbol{x},t) u_{\alpha}(\boldsymbol{x},t)\rangle},
\end{align}
where $r$ denotes the distance between the evaluated fluid velocities and $\boldsymbol{e}_{\parallel}$ and $\boldsymbol{e}_{\parallel}$ the unit vector in the longitudinal and transverse direction, respectively. Note that no summation is carried out over the index $\alpha$. \\
It can be seen in figure \ref{fig:autcorrelations}, that the fluid velocity decorrelates faster in the transverse direction than in the longitudinal direction, which is qualitatively also captured by the two LES. The LES without wavelet enrichment, however, predicts too slow decorrelation of the velocity in the longitudinal and transverse direction. The decorrelation can be accelerated by the wavelet enriched LES, as it provides the unresolved subgrid-scale velocity field. As a results, the autocorrelations of the wavelet enriched LES are in very good agreement with the autocorrelations of the DNS for small distances $r$. As the distance increases, the energetic resolved eddies dominate the autocorrelation and the autocorrelations of the wavelet enriched LES converge towards the autocorrelations of the LES without wavelet enrichment. \\ 
 Figure \ref{fig:gradients} shows the PDF of the longitudinal and transverse velocity gradients, $A_{11}=\partial u_1/\partial x_1$ and $A_{12}=\partial u_1/\partial x_2$, normalized by their respective standard deviations, $\sigma_{11}$ and $\sigma_{12}$. Similar to the DNS, the wavelet enriched LES increases the probability of events of high magnitude of the gradients compared to the LES. The tails of the PDFs of the wavelet enriched LES are wider than in a Gaussian distribution, which is also observed in the DNS. In the PDF of the transverse velocity gradients, the wavelet enriched LES shows a good agreement with the shape of the DNS. For the longitudinal gradients, the PDF of the wavelet enriched LES lacks asymmetry, but the overall agreement with the PDF of the DNS is significantly improved compared to the LES without wavelet enrichment. \\
The analysis of the forced HIT proves that characteristic properties of turbulence can be reproduced by the wavelet enrichment, which can not be achieved by, for instance, sampling random coefficient of Fourier-modes as done by a kinematic simulation (see, e.g., \citet{Zhou2020b}). \\
The computational cost of the wavelet enriched LES for the simulation configuration HIT-f-s are approximately a factor 3-4 of the computational cost of the LES without enrichment. The computational cost of the corresponding DNS is four orders of magnitude higher. \\
Even though the wavelet enrichment proves to be capable of predicting realistic single-phase flow statistics in HIT, the vast majority of relevant flows in industry and nature is inhomogeneous and anisotropic. A simple inhomogeneous and anisotropic flow configuration is the turbulent shear flow (TS-s) introduced in section \ref{sec:simulations}. In figure \ref{fig:shearflowcorrelations}, we compare the spatial correlations of the subgrid-scale velocity, which are defined as
\begin{align}
    \mathcal{C}_{ij} = \langle (u_i^{\prime} - \langle u_i^{\prime} \rangle) (u_j^{\prime} - \langle u_j^{\prime} \rangle)\rangle,
\end{align}
where $\langle .\rangle$ indicates temporal averaging and averaging along the spatially homogeneous directions (i.e., x- and z-directions). The reference subgrid-scale fluid velocity field is obtained by subtracting the explicitly filtered fluid velocity field of the DNS from the unfiltered fluid velocity field of the DNS, where a spectrally sharp filter is used. Because of the arbitrary choice of the filter, we compare the results only qualitatively. \\
It can be seen from figure \ref{fig:shearflowcorrelations} that the correlations of the wavelet enrichment are not as smooth as the correlations of the DNS. Because of the shape of the wavelet basis functions, regions occur that have statistically higher kinetic energy than other regions. Owing the temporarily changing positions of the wavelet basis functions, which is explained in section \ref{ssec:realization}, the fluctuations are significantly reduced, but fluctuations remain in the ensemble averaged correlations that, however, are much smaller than the instantaneous velocity fluctuations. \\
The velocity correlations $\mathcal{C}_{11}$, $\mathcal{C}_{22}$, and $\mathcal{C}_{33}$ vary with two periods of a sine-shape across the y-direction. With the wavelet enrichment this shape can be reproduced, albeit with different magnitudes, which can be traced back to the filter choice of the DNS and the estimation of the subgrid-scale kinetic energy. The cross-correlation $\mathcal{C}_{12}$ varies with one period of a sine-shape. At the domain center the velocity in x- and y- direction are positively correlated and at the domain boundary negatively correlated. The wavelet enrichment reproduces this trend, which proves that the right anisotropic behavior can be generated from information that the wavelet enrichment receives from the LES. \\

\subsection{Particle-laden flow statistics}
\begin{figure}[hp]
    \centering
    \includegraphics[scale=0.9]{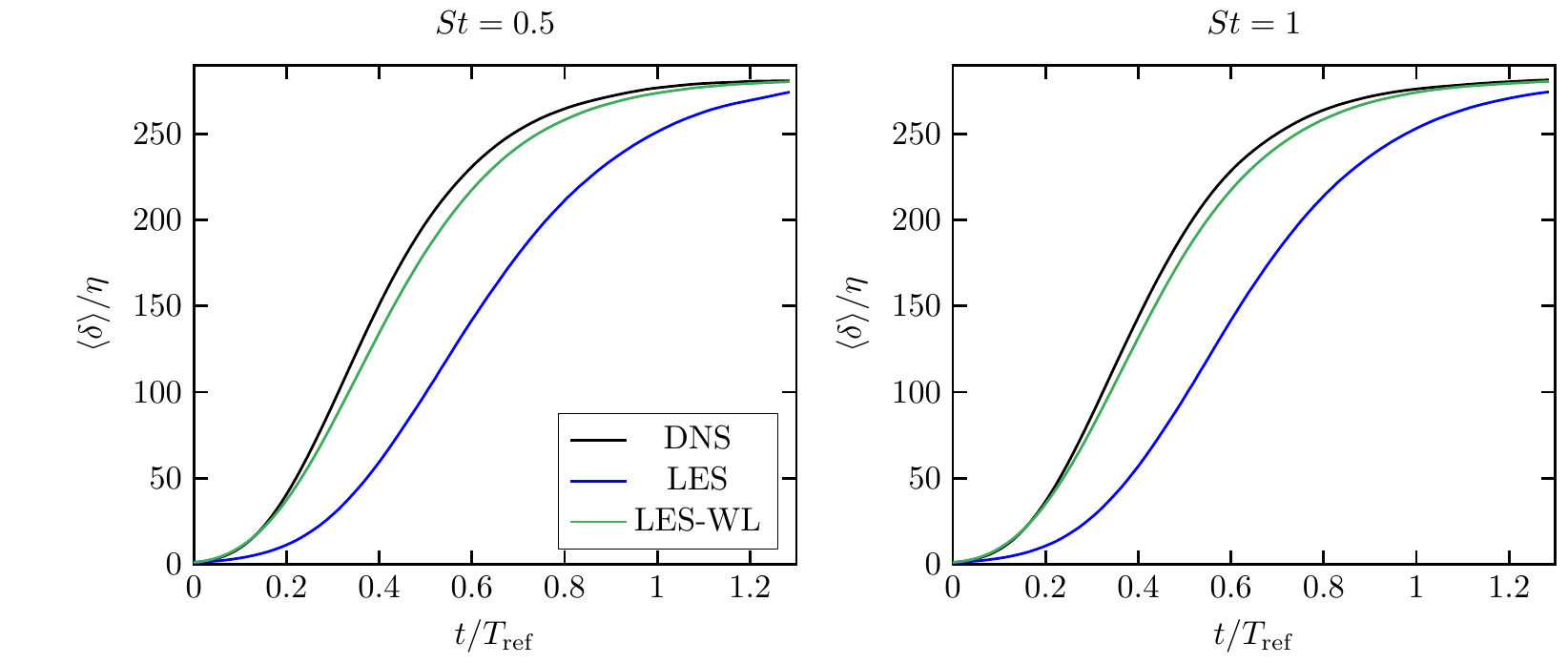}
    \includegraphics[scale=0.9]{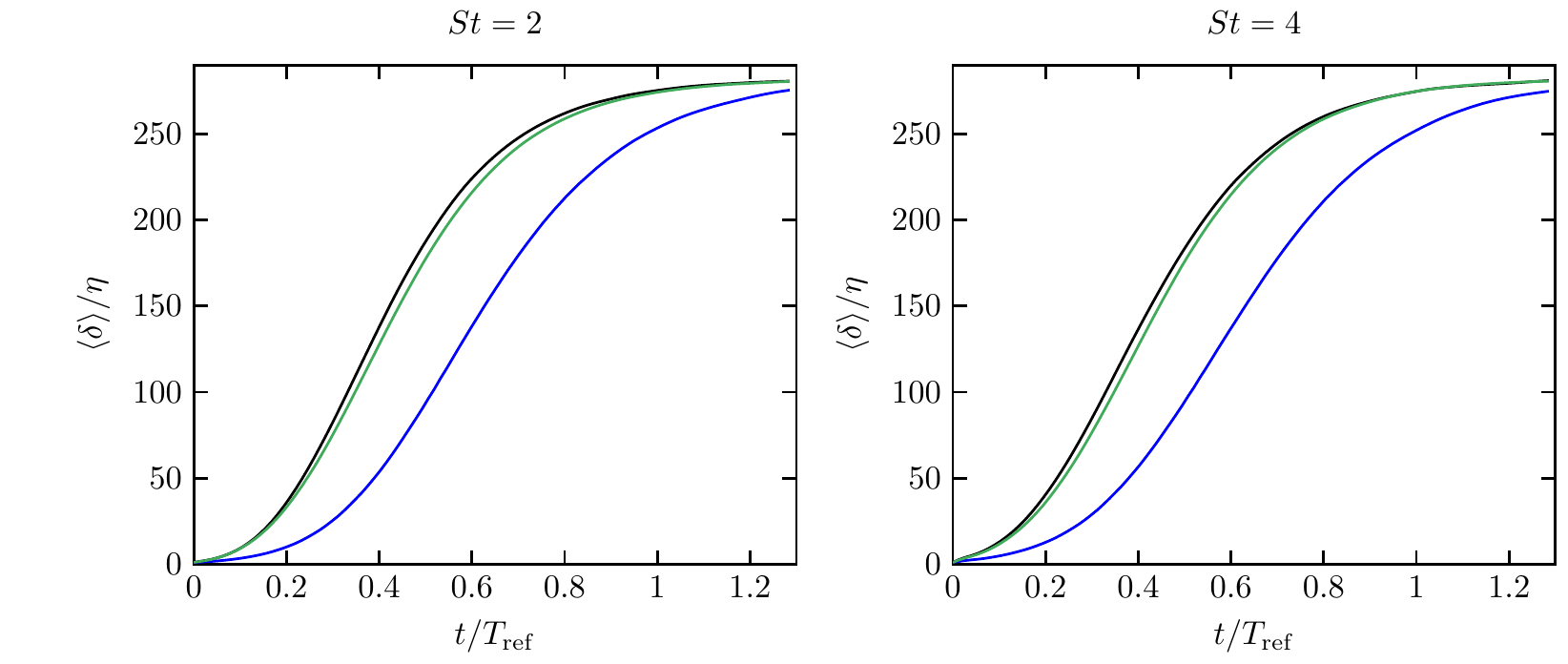}
    \includegraphics[scale=0.9]{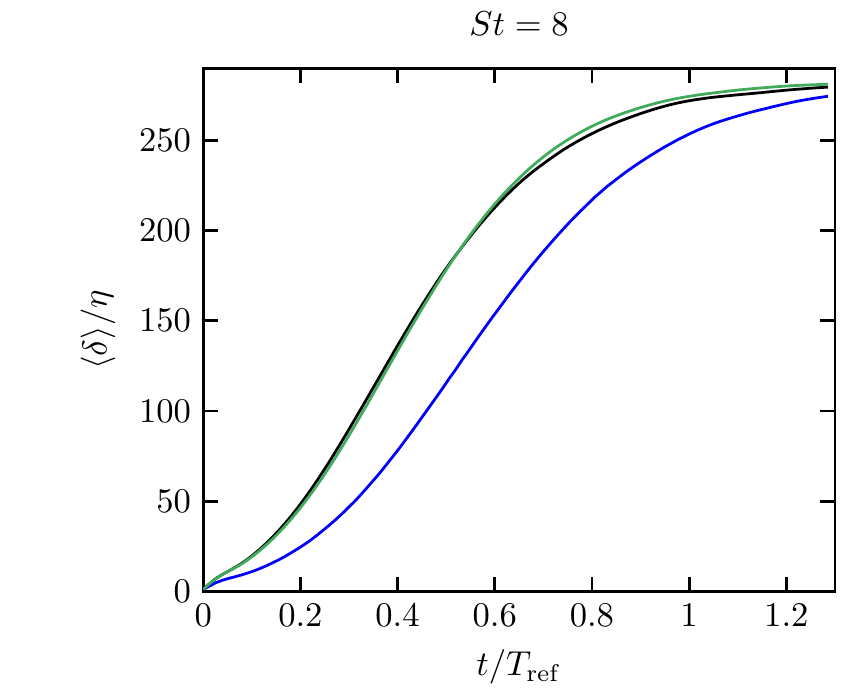}
    \caption{Particle pair dispersion of particles with Stokes numbers in the range $St\in\{0{.}5,1,2,4,8\}$ for the simulation configuration HIT-f-1wc. The considered particles pairs have an initial separation of the size of the Kolmogorov length scale $\eta$. The results are shown for the DNS, the LES, and the wavelet enriched LES (LES-WL).}
    \label{fig:pps}
\end{figure}

\begin{figure}[hp]
    \centering
    \includegraphics[scale=0.9]{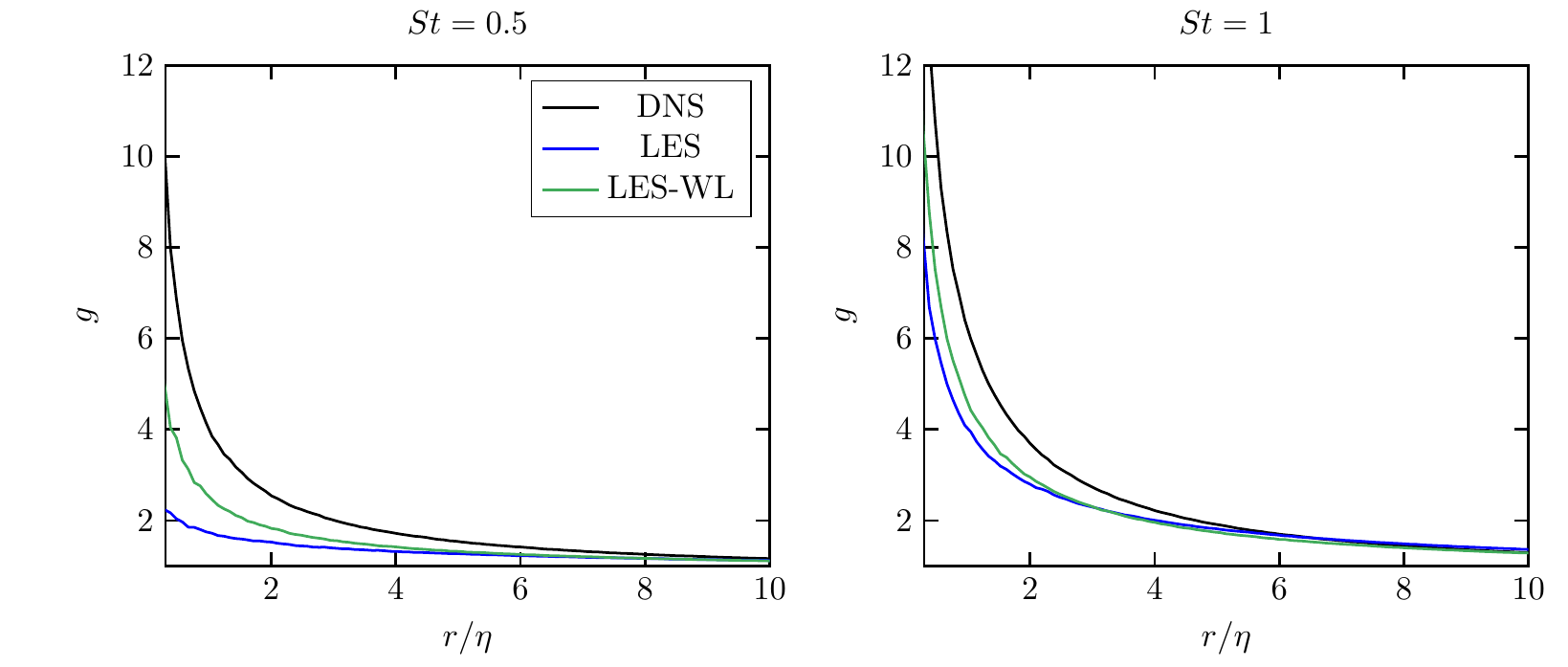}
    \includegraphics[scale=0.9]{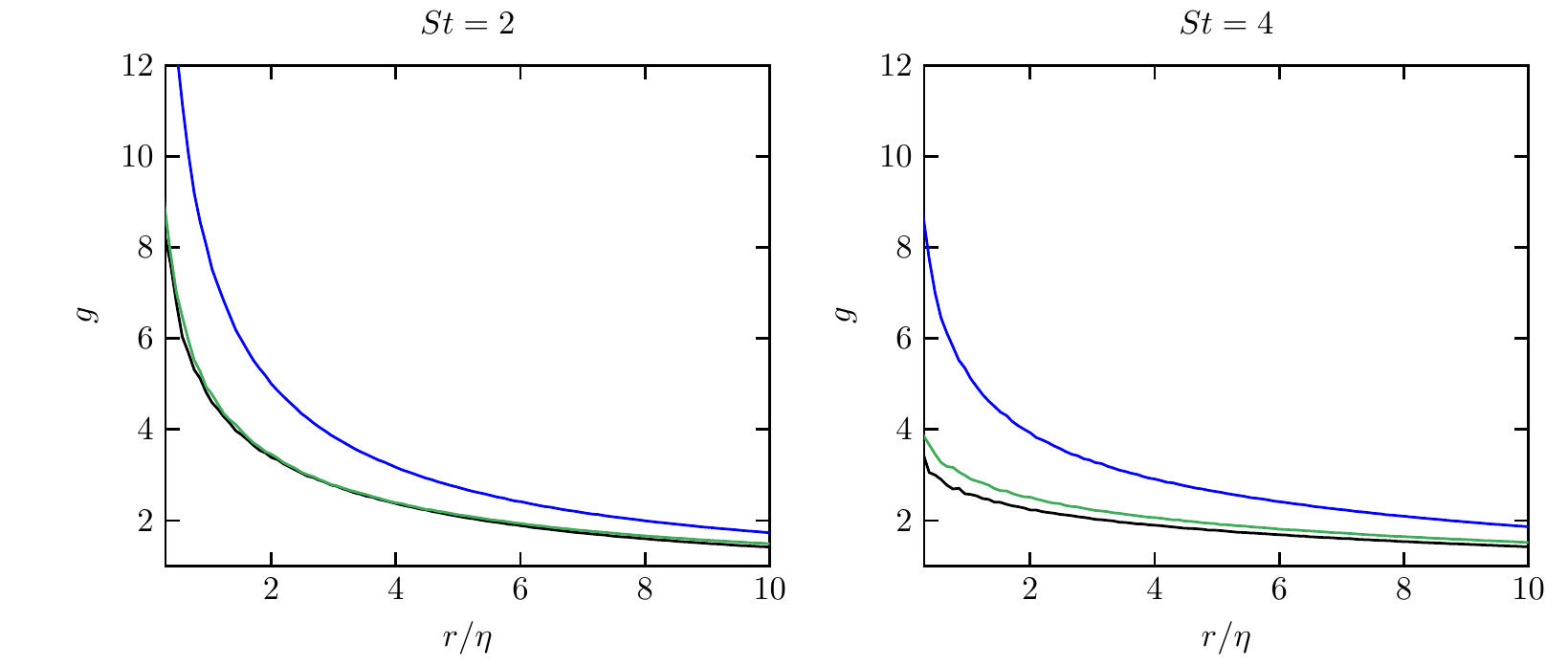}
    \includegraphics[scale=0.9]{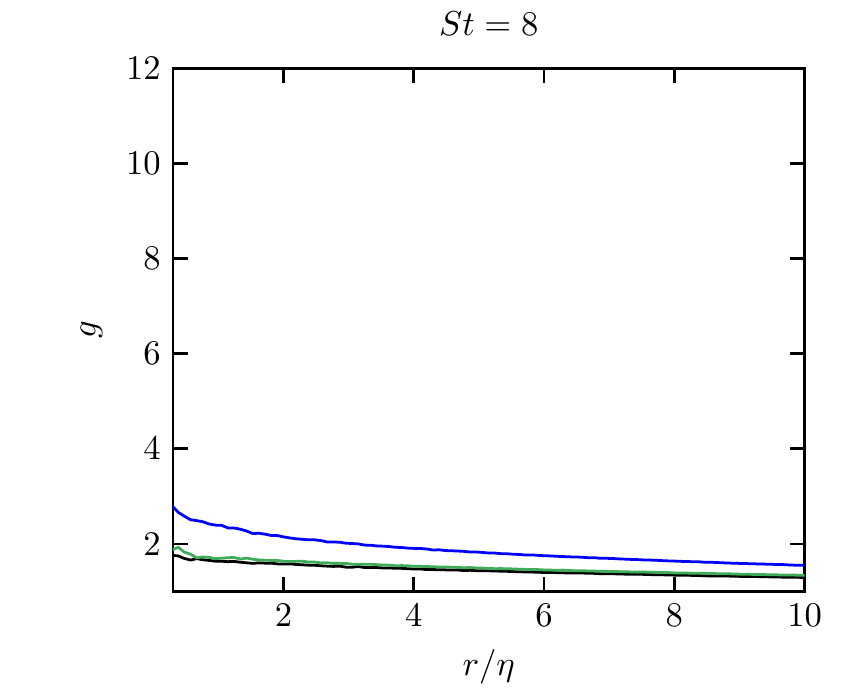}
    \caption{Radial distribution function of particles with Stokes numbers in the range $St\in\{0{.}5,1,2,4,8\}$ for the simulation configuration HIT-f-1wc. The results are shown for the DNS, the LES, and the wavelet enriched LES (LES-WL).}
    \label{fig:rdf}
\end{figure}

\begin{figure}[h]
    \centering
    \subfigure[]{\label{fig:twowaya}\includegraphics[scale=0.9]{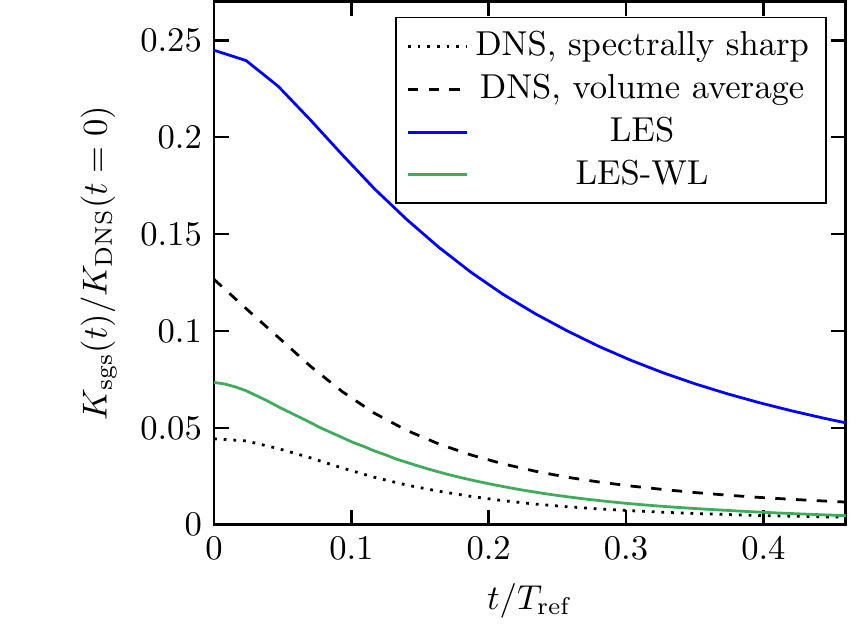}}
    \subfigure[]{\label{fig:twowayb}\includegraphics[scale=0.9]{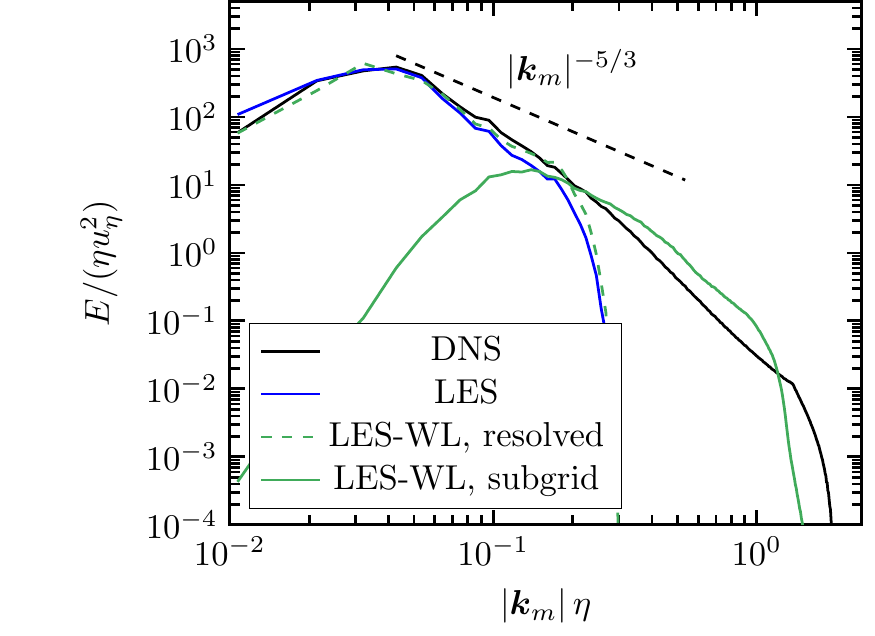}}
    \caption{Subgrid-scale kinetic energy over time (a) and kinetic energy spectrum before the onset of the decay (b) in decaying HIT laden with particles of Stokes number $St=8$ for the simulation configuration HIT-d-2wc. The results are shown for the DNS, the LES, and the wavelet enriched LES (LES-WL). The inertial range slope is plotted for comparison.}
    \label{fig:twoway}
\end{figure}

We evaluate the particle statistics by comparing the results of forced HIT laden with one-way coupled particles (HIT-f-1wc). Figure \ref{fig:pps} shows the particle pair dispersion, which is defined as the ensemble averaged temporal evolution of the distance between particle pairs with the position $\boldsymbol{x}_{p0}(t)$ and $\boldsymbol{x}_{p1}(t)$:
\begin{align}
    \langle \delta \rangle(t) = \langle |\boldsymbol{x}_{p0}(t) - \boldsymbol{x}_{p1}(t)| \rangle.
\end{align}
A particle pair is defined as two particles that have an initial separation of approximately the Kolmogorov length scale. \\
For all the considered Stokes numbers, three phases of the dispersion are observed in figure \ref{fig:pps}: (\romannumeral 1) particle pairs are located in the same eddy and stay close together, before (\romannumeral 2), they rapidly disperse by experiencing widely uncorrelated fluid velocities, and (\romannumeral 3), their maximum separation is reached, which is determined by the domain size. Since the LES only contains the largest eddies, that even the particles with the largest considered Stokes number can follow well, the dispersion is much slower than for the particles transported with the DNS velocity field. The wavelet enriched LES leads to a particle pair dispersion that almost coincides with the DNS, for all considered Stokes numbers. \\
Most important for many applications is a correct prediction of particle clustering, i.e., that the particles preferentially concentrate in specific regions. A quantitative measure of particle clustering is provided by the radial distribution function, defined as
\begin{align}
    g(r) = \left\langle \dfrac{N_{\mathrm{p},i}(r)/\Delta V_i(r)}{N_{\mathrm{p}}/V} \right\rangle,
\end{align}
where $\Delta V_i(r)$ is the volume of spherical shells in a distance $r$, and  $N_{\mathrm{p},i}(r)$ is the number of particle in the respective spherical shell. The radial distribution function is normalized with the total number of particles $N_{\mathrm{p},i}(r)$ and the total volume of the simulation domain $V$. \\
The radial distribution function in figure \ref{fig:rdf} predicted by the LES without wavelet enrichment significantly deviates from the radial distribution function predicted by the DNS, whereas the LES predicts too little clustering for $St\in\{0{.}5, 1\}$ and too strong clustering for $St\in\{2,4,8\}$. For the small Stokes numbers, the small eddies, that are absent in the LES without wavelet enrichment, move the particles towards regions of small vorticity and large strain. The particles with the larger Stokes numbers can not follow the small eddies well. Consequently, the small fluid velocity structures increase the dispersion of the particles with large Stokes numbers. The correct prediction of clustering of the particles with small Stokes numbers is far more challenging, because the strain-rotation relations of the small fluid velocity structures has a higher impact. It is observed in figure \ref{fig:rdf} that the wavelet enriched LES significantly improves the radial distribution function for the Stokes numbers $St\ge2$. The increased dispersion by the modeled subgrid-scale fluid velocity field yields an excellent agreement of the radial distribution function of the wavelet enriched LES with the radial distribution function of the DNS.
For the intricate case of Stokes numbers $St=0{.}5$ and $St=1$, only a relatively minor increase in particle clustering is observed.\\
The final test case considered in this paper considers two-way coupling. The abilities of the wavelet enrichment to predict the unresolved effects of two-way coupled particle-laden flows is assessed with the simulation configuration HIT-d-2wc, representing decaying HIT with two-way coupled particles with Stokes number $St=8$. Figure \ref{fig:twowaya} shows the temporal evolution of the subgrid-scale kinetic energy predicted by the LES without the wavelet enrichment and the LDKM, the wavelet enriched LES with the mLDKM including the effect of the particles, and the DNS.

The reference subgrid-scale kinetic energy is obtained by explicit filtering the DNS. In a LES, the spatially varying turbulent viscosity imposes the filtering. This unknown filter leads to an uncertainty of the actual subgrid-scale kinetic energy. Therefore, the DNS is explicitly filtered with two different filters, a spectrally sharp filter and by volume averaging. During the whole decay the LES without wavelet enrichment and with the single-phase flow subgrid-scale model predicts a much too large subgrid-scale kinetic energy. The subgrid-scale model, i.e., the LDKM, assumes single-phase turbulence for the unresolved velocity scales, which contains more kinetic energy than the same flow laden with particles of Stokes number $St=8$. Since the wavelet enriched LES enables the use of the mLDKM that takes the turbulence modification by the particles into account, the predicted subgrid-scale kinetic energy is much smaller and lies between the subgrid-scale kinetic energy obtained from the DNS using the two different explicit filters. Note that the subgrid-scale kinetic energy obtained from the DNS is sensitive to the specific filter that is used. \\
The kinetic energy spectra before the onset of the decay are shown in figure \ref{fig:twowayb}. The turbulence modulation by the particles leads to a clear deviation of the DNS spectrum from the well known power-law of the inertial range that is observed in single-phase turbulence. Close to the cutoff wave number, the spectrum of the LES without wavelet enrichment using the LDKM deviates from the DNS spectrum because of the overestimated subgrid-scale kinetic energy and, thus, the too large turbulent viscosity. This is improved by the wavelet enriched LES using the mLDKM that predicts a smaller turbulent viscosity and a better agreement with the kinetic energy spectrum of the DNS. The kinetic energy of the subgrid-scale velocity generated by the wavelet enrichment is slightly larger than the kinetic energy of the DNS. The shape of the kinetic energy spectrum, however, is very similar to the kinetic energy spectrum of the DNS.
\section{Conclusions}
\label{sec:conclusions}
We propose a novel model to predict the unresolved subgrid-scale fluid velocity field in the scope of a LES. In LES of particle-laden turbulent flows, the subgrid-scale velocity at the particle positions is required to predict correct particle behavior, such as their dispersion and clustering. The new wavelet enrichment model discretizes the subgrid-scale velocity by means of a divergence-free wavelet vector basis. The coefficients of this basis are obtained by minimizing the squared error of the linearized SFNSE. In contrast to structural models using a Fourier basis, the novel wavelet enrichment enables a continuous change of velocity statistics across the domain and, hence, the generation of an inhomogeneous subgrid-scale velocity field. Furthermore, the wavelet enrichment does not require to specify parameters that critically affect the results.  \\
The novel model is validated with 4 distinct test cases that separately assesses the predictions of the wavelet enrichment in single-phase and particle-laden flows. Simulations of forced HIT show that the wavelet enrichment produces strain-rotation relations that are similar to the DNS. The PDFs of the longitudinal and transverse velocity gradients possess the expected non-Gaussian behavior. The wavelet enrichment is shown to be capable of predicting inhomogeneous and anisotropic velocity fields in a turbulent shear flow, where the spatial velocity correlations match those of DNS. \\
One-way coupled simulations of forced HIT of particles with different Stokes numbers revealed excellent agreement with the particle pair dispersion of the DNS. The predictions of particle clustering are improved, whereas the wavelet enriched LES recovers the radial distribution function of the DNS very well for Stokes numbers $St\ge2$. By combining the wavelet enrichment with the recently proposed modification of the LDKM \cite{Hausmann2023} to particle-laden flows, we report improved predictions of the subgrid-scale kinetic energy and the kinetic energy spectrum in two-way coupled decaying HIT. \\
The proposed wavelet enrichment is able to recover the most important interactions between turbulence and particles while maintaining computational costs of the order of the costs of a LES.  

\begin{acknowledgments}
This research was funded by the Deutsche Forschungsgemeinschaft (DFG, German Research Foundation) \textemdash Project-ID 457509672. 
\end{acknowledgments}

%

\appendix
\section{Movement of the wavelets}
\label{sec:appendix}
Let $\boldsymbol{x}_{j,\boldsymbol{k},\mathrm{c}}(t)$ be the coordinates of the center positions of the wavelets with the indices $j$ and $\boldsymbol{k}$ that are initially given as
\begin{align}
    x_{i,j,\boldsymbol{k},\mathrm{c}}(t=0) = \dfrac{x_{i,\mathrm{max}} - x_{i,\mathrm{min}}}{2^j}(k_i+1/2)+x_{i,\mathrm{min}}.
\end{align}
A wavelet center is allowed to move within a region
\begin{align}
    \boldsymbol{x}_{j,\boldsymbol{k},\mathrm{c}}(t=0)-\boldsymbol{\delta}_j \leq \boldsymbol{x}_{j,\boldsymbol{k},\mathrm{c}}(t) \leq  \boldsymbol{x}_{j,\boldsymbol{k},\mathrm{c}}(t=0)+\boldsymbol{\delta}_j,
\end{align}
where 
\begin{align}
    \delta_{i,j} = \dfrac{x_{i,\mathrm{max}} - x_{i,\mathrm{min}}}{2^{j+1}}.
\end{align}
Within this region target coordinates $\boldsymbol{x}_{j,\boldsymbol{k},\mathrm{t}}$ are generated from a uniform distribution. The wavelet center moves towards the target coordinates with a constant velocity that is equal to $10\%$ of the local eddy turnover time. As soon as the center coordinates are close to the target coordinates, new random target coordinates are generated.

\end{document}